\documentstyle[epsfig]{ioplppt}          % use this for journal style
\newcommand{\JETSET}{\mbox{\sc JETSET}}
\newcommand{\PYTHIA}{\mbox{\sc PYTHIA}}
\newcommand{\HERWIG}{\mbox{\sc HERWIG}}
\newcommand{\ARIADNE}{\mbox{\sc ARIADNE}}
\newcommand{\EXCALIBUR}{\mbox{\sc EXCALIBUR}}
\newcommand{\WPHACT}{\mbox{\sc WPHACT}}
\newcommand{\KORALW}{\mbox{\sc KORALW}}
\newcommand{\LEPONE}{\mbox{\sc LEP 1}}
\newcommand{\LEPTWO}{\mbox{\sc LEP 2}}
\newcommand{\QCD}{\mbox{\sc QCD}}
\newcommand{\ISR}{\mbox{\sc ISR}}
\newcommand{\BW}{\mbox{\sc BW}}
\newcommand{\CERN}{\mbox{\sc CERN}}
\newcommand{\NMR}{\mbox{\sc NMR}}
\def\gev{\;{\rm GeV}}
\def\mev{\;{\rm MeV}}
\def\pb{\;{\rm pb}}
\def\lapproxeq {\mbox{{\lower .7ex\hbox{$\;\stackrel{\textstyle
                  <}{\sim}\;$}}}}
\def\gapproxeq  {\mbox{{\lower .7ex\hbox{$\;\stackrel{\textstyle
                  >}{\sim}\;$}}}}
\def\cpc#1#2#3{19#3 {\em Comp.\ Phys.\ Commun.}~{\bf#1} #2}
\def\hepph#1#2{19#1 \verb"hep-ph/#1#2"}

\def\pl#1#2#3{19#3 {\em Phys.\ Lett.}~{\bf#1B} #2}
\def\pr#1#2#3{19#3 {\em Phys.\ Rev.}~D {\bf#1} #2}

\def\zp#1#2#3{19#3 {\em Z.\ Phys.}~C {\bf#1} #2}
\newcommand{\nchqqqq}{\mbox{$N_{\mathrm{ch}}^{\mathrm{4q}}$}}
\newcommand{\nchqqlv}{\mbox{$N_{\mathrm{ch}}^{\mathrm{qq\ell\nu}}$}}
\newcommand{\mnchqqqq}{\mbox{$\langle\nchqqqq\rangle$}}
\newcommand{\mnchqqlv}{\mbox{$\langle\nchqqlv\rangle$}}
\newcommand{\WW}{\mbox{$\mathrm{W^+W^-}$}}
\newcommand{\qq}{\mbox{$\mathrm{q\overline{q}}$}}

\newcommand{\lv}{\mbox{$\ell\overline{\nu}_{\ell}$}}
\newcommand{\WWqqqq}{\mbox{\WW$\rightarrow$\qq\qq}}
\newcommand{\WWqqlv}{\mbox{\WW$\rightarrow$\qq\lv}}
\newcommand{\MW}{\mbox{$M_{\mathrm{W}}$}}
\newcommand{\Zz}{\mbox{${\mathrm{Z}^0}$}}
\newcommand{\Zqq}{\mbox{$\Zz/\gamma\rightarrow\qq$}}
\newcommand{\delnch}{\mbox{$\Delta\langle\nchqqqq\rangle$}}
\newcommand{\dndy}{\mbox{${\rm d}N_{ch}/{\rm d}y$}}
\newcommand{\epem}{\mbox{$\mathrm{e^+e^-}$}}
\newcommand{\kt}{\mbox{$k_{\perp}$}}
\newcommand{\ee}{\mbox{${\mathrm{e}}^+{\mathrm{e}}^-$}}
\newcommand\new{\newcommand}         % shorthand for \newcommand
\catcode`\"=\active                  % change category for \"
\new"[1]{{\accent127#1}}             % permit to do umlauts as "a
\new\3{\ss }                         % replace \ss by \3
\new{\mm}[1]{{\mbox{\hspace{#1mm}}}} % create horizontal space

\new\Tab[1]{Table~\ref{tab:#1}}
\newcommand {\GW}      {\Gamma_{\mathrm{W}}}
\newcommand {\MTRUE}      {M_{\mathrm{true}}}
\newcommand {\MREC}      {M_{\mathrm{rec}}}
\newcommand {\MFIT}      {M_{\mathrm{fit}}}
\newcommand {\aaa}       {\it{a}}
\newcommand {\bbb}       {\it{b}}
\newcommand {\MLNREC}      {m(\ell\nu)_{\mathrm{rec}}}
\newcommand {\qqqqa}       {\rm{\qqb\qbq}}
\newcommand {\qqen}       {\rm{\qqb e}\nbel}
\newcommand {\qqmn}       {\rm{\qqb}\mu\nbmu}
\newcommand {\qqtn}       {\rm{\qqb}\tau\nbtau}
\newcommand {\qqln}       {\rm{\qqb}\ell\nbl}
\newcommand {\uden}       {\rm{e}\overline{\nu}\rm{u\overline{d}}}
\newcommand {\udmn}       {\mu\overline{\nu}\rm{u\overline{d}}}

\newcommand {\Wjjln}       {{\rm WW}\rightarrow {\rm{jj}\ell}\nu}
\newcommand {\Wjjtn}       {{\rm WW}\rightarrow {\rm{jj}\tau}\nu}

\newcommand {\Wjjjj}        {{\rm WW}\rightarrow {\rm {jjjj}}}

\new\dmw{\mbox{$\Delta \MW$}}
\newcommand {\chit}       {\chi^{2}}
\newcommand{\ALEPH}{\mbox{\sc ALEPH}}
\newcommand{\DELPHI}{\mbox{\sc DELPHI}}
\newcommand{\OPAL}{\mbox{\sc OPAL}}
\newcommand{\LTHREE}{\mbox{\sc L3}}
\newcommand{\LEP}{\mbox{\sc LEP}}
\newcommand{\CCTHREE}{\mbox{\tt CC03}}
\newcommand{\nbel}{{{\overline{\nu}}_{e}}}
\newcommand{\en}{{e \nbel}}
\newcommand{\qqb}{{\mathrm{q{\bar q'}}}}
\newcommand{\enqq}{{\mathrm{ \en  \qqb}}}
\newcommand{\nl}{{\nu_{\ell}}}
\newcommand{\nbl}{{{\overline{\nu}}_{\ell}}}
\newcommand{\lnln}{{\ell \nbl {\overline{\ell}} \nl}}
\newcommand{\nbmu}{{{\overline{\nu}}_{\mu}}}
\newcommand{\nbtau}{{{\overline{\nu}}_{\tau}}}
\newcommand{\mn}{{\mu \nbmu}}
\newcommand{\mnqq}{{\mathrm{ \mn  \qqb}}}
\newcommand{\qbq}{{{\bar q}q'}}
\newcommand{\qqqq}{\mbox{$\mathrm{ \qqb \qbq}$}}
\newcommand{\len}{{\ell \nbl}} 
\newcommand{\lnqq}{{\mathrm{ \len  \qqb}}}
\begin{document} 

\title{Report of the Working Group on `W Mass and QCD'}[W Mass and QCD]
 
\author{A~Ballestrero$^1$, D~G~Charlton$^2$, G~Cowan$^3$, P~Dornan$^4$, 
R~Edgecock$^5$, J~Ellis$^6$, E~W~N~Glover$^7$, C~Hawkes$^8$,
H~Hwang$^9$, R~Jones$^{10}$, V~Kartvelishvili$^{11}$, Z~Kunszt$^{12}$, 
E~Maina$^1$, D~J~Miller$^{5}$, S~Moretti$^8$, A~Moutoussi$^4$, 
C~Parkes$^9$,  P~B~Renton$^{9\ast}$, D~A~Ross$^{13}$, W~J~Stirling$^{7\ast}$,
J~C~Thompson$^5$, M~Thomson$^6$, D~R~Ward$^{8\ast}$, C~P~Ward$^8$, 
J~Ward$^{14}$, M~F~Watson$^6$, 
N~K~Watson$^2$, B~R~Webber$^8$} 
 
\address{
$^1$University of Torino, Italy \\ 
$^2$University of Birmingham, UK \\ 
$^3$University of Siegen, Germany \\ 
$^4$ICSTM, London, UK \\ 
$^5$Rutherford Appleton Laboratory, UK \\ 
$^6$CERN, Geneva, Switzerland \\ 
$^7$University of Durham, UK \\ 
$^8$University of Cambridge, UK \\ 
$^9$University of Oxford, UK \\ 
$^{10}$University of Lancaster, UK \\ 
$^{11}$University of Manchester, UK \\ 
$^{12}$ETH Zurich, Switzerland \\ 
$^{13}$University of Southampton, UK \\ 
$^{14}$University of Glasgow, UK \\ 
\\ 
$^\ast$\ convenors}

\begin{abstract} 
The W Mass and QCD Working Group discussed a wide variety of topics 
relating to present and future measurements of $M_W$ at \LEPTWO, including 
\QCD\ backgrounds to $W^+W^-$ production. Particular attention was focused on
experimental issues concerning the direct reconstruction and threshold
mass measurements, and on theoretical and experimental 
issues concerning the four jet final state. This report summarises
the main conclusions.
\end{abstract} 
 
% 
%  Uncomment out if preprint format required 
% 
%\pacs{00.00, 20.00, 42.10} 
%\maketitle

\setcounter{footnote}{1}
\section{Introduction\/\protect\footnotemark[\value{footnote}]}
\footnotetext[\value{footnote}]{Unless otherwise stated, the sections 
have been prepared by the convenors.}

The `W Mass and QCD Working Group' addressed a variety of topical 
questions
during the Workshop. The format varied from formal seminar
presentations to informal discussions, with a  total of 28 people
contributing. In this article we summarise the outcome of the
discussions, including in particular new results obtained both
during and after the meeting. We do not attempt to review the 
status of the various physics topics prior to the meeting,
as this was very well covered in the plenary talks
by J C Thompson \cite{JT}, B R Webber \cite{BW} and G Cowan \cite{GC}. 

For most of the time the Working Group separated into two partially
overlapping subgroups. The first focused on theoretical and experimental
issues concerning various aspects of the final state in $W^+W^-$ and
\QCD\ four jet production, in particular colour reconnection,
Bose-Einstein correlations, and the accuracy of current \QCD\ models
for the four jet final state. The second subgroup was concerned with
mainly experimental issues concerning the direct reconstruction
and threshold cross section methods for measuring $M_W$ at \LEPTWO.
In addition, the subgroup updated the expected precision of the two 
methods based on experience with the two methods to date.

The report is organised as follows. The work of the two subgroups
is described in Sections \ref{sub1} and \ref{sub2}. Each section contains
a general overview followed by individual contributions as subsections.
The overall conclusions of the Working Group are presented in Section
\ref{conc}.

%\setcounter{footnote}{1}
%\section{Aspects of the hadronic final state 
%in $W^+W^-$ production\/\protect\footnotemark[\value{footnote}]}
%\footnotetext[\value{footnote}]{Prepared by ... }
%\label{sub1}
\section{Aspects of the hadronic final state 
in $W^+W^-$ production}
\label{sub1}

\setcounter{footnote}{3}
\subsection{Experimental aspects of colour reconnection\/
\protect\footnotemark[\value{footnote}]}
\footnotetext[\value{footnote}]{Prepared by M~F~Watson, N~K~Watson}

\label{colrec}

Colour reconnection (also referred to as `rearrangement' or `recoupling')
in \WW\ decays has been the subject of many studies (e.g.\ 
\cite{GPZ,SK,YB}) and at present there is agreement that observable effects
of interference between the colour singlets in the perturbative phase are
expected to be small.  In contrast, significant interference in the
hadronisation process appears a viable prospect but, with our current lack
of knowledge of non-perturbative \QCD, such interference can only be
estimated in the context of specific models
\cite{SK,GH,ARIADNE,HERWIG,NOVA,EG}. In the studies described below,
experimentally accessible features of these models\footnote{In studying
  these models, no retuning was performed when reconnection was enabled.}
are investigated, paying particular attention to the bias introduced to a
typical measurement of \MW\ by direct reconstruction of the decay products.

Throughout this section reconnection effects were studied using:
\PYTHIA\ 5.7~\cite{tt11},
%\PYTHIA~5.722, 
type I and type II superconductor models (with the string
overlap integral in the type I case characterised by $\rho=0.9$) \cite{SK,YB};
\ARIADNE~4.08 allowing reconnection between the two W bosons; and
\HERWIG~5.9, in both its default reconnection model and also a `colour
octet' variant in which merging of partons to form clusters was performed on
a nearest neighbour basis\footnote{This was suggested by B~R~Webber, as a
  partial emulation of the model of reference \cite{EG}.}. In all cases, the
tuning of the models was as used in reference \cite{OPAL161}.

\subsubsection{Inclusive charged multiplicity}

It has been suggested \cite{SK,GH} that simple observable quantities such
as the charged multiplicity in restricted rapidity intervals may be
sensitive to the effects of colour reconnection. More recently \cite{EG} it
was suggested that the effect on the inclusive charged multiplicity itself
may be larger than previously considered and that the mean hadronic
multiplicity in \WWqqqq\ events, \mnchqqqq, may be as much as 10\% smaller
than twice the hadronic multiplicity in \WWqqlv\ events, \mnchqqlv. It was
also reported during this workshop that the effects of Bose-Einstein
correlations may increase \mnchqqqq\ by $\sim 3$--$10$\% 
(see section~\ref{bose-eins}).

The shifts in \mnchqqqq\ at the hadron level predicted by the models
studied thus far are given in table~\ref{tab:cr_nch}, where \delnch\ is
defined as the change in mean multiplicity relative to the `no
reconnection' scenario of each model.  From these, it is clear that the
multiplicities themselves and also the magnitude and sign of the predicted
shifts are model dependent.

\begin{table}
\caption{Mean charged multiplicities, \mnchqqqq,
 and predicted shifts for various models}
 \label{tab:cr_nch}
\begin{indented}
 \lineup
\item[]\begin{tabular}{@{}llll}
\br
model & & \mnchqqqq\ & \protect\delnch\ (\%) \\
\mr
 \PYTHIA\  &  normal      &  38.64 & \\
         &  type I      &  38.21 & $-1.1\pm$0.1 \\
         &  type II     &  38.39 & $-0.7\pm$0.1 \\
 \HERWIG\  &  normal      &  37.07 & \\
         &  reconnected ($P=\frac{1}{9}$) &  37.25  & +0.5$\pm$0.1 \\
         &  reconnected ($P=1$) &  38.38  & +3.5$\pm$0.1 \\
 \ARIADNE\ &  normal      &  38.14 &               \\
         &  reconnected &  37.07 &         $-2.8\pm$0.1 \\
\br
\end{tabular}
\end{indented}
\end{table}

In this study, the precision with which such tests may be performed is
quantified.  As a starting point for such tests, it was first verified that
in the absence of reconnection effects $\mnchqqqq=2\mnchqqlv$ in the models
\PYTHIA\ and \HERWIG. The statistical uncertainty of this test was ${\cal
  O}(0.1\%)$.  Next, samples of $10^5$ \HERWIG\ and \PYTHIA\ \WW\ events were
generated at $\sqrt{s}=171$~GeV with a full simulation of the \OPAL\
detector, and realistic event selections were applied for both \WWqqqq\ and
\WWqqlv\ ($\ell=$ e, $\mu$ and $\tau$).  The efficiency in each case was
$\sim$80\%, while the purity is $\sim 80\%$ for \WWqqqq\ and $\sim 88\%$
for the \WWqqlv\ channel.

\begin{figure}[tb]
  \epsfxsize=\textwidth
  \epsffile{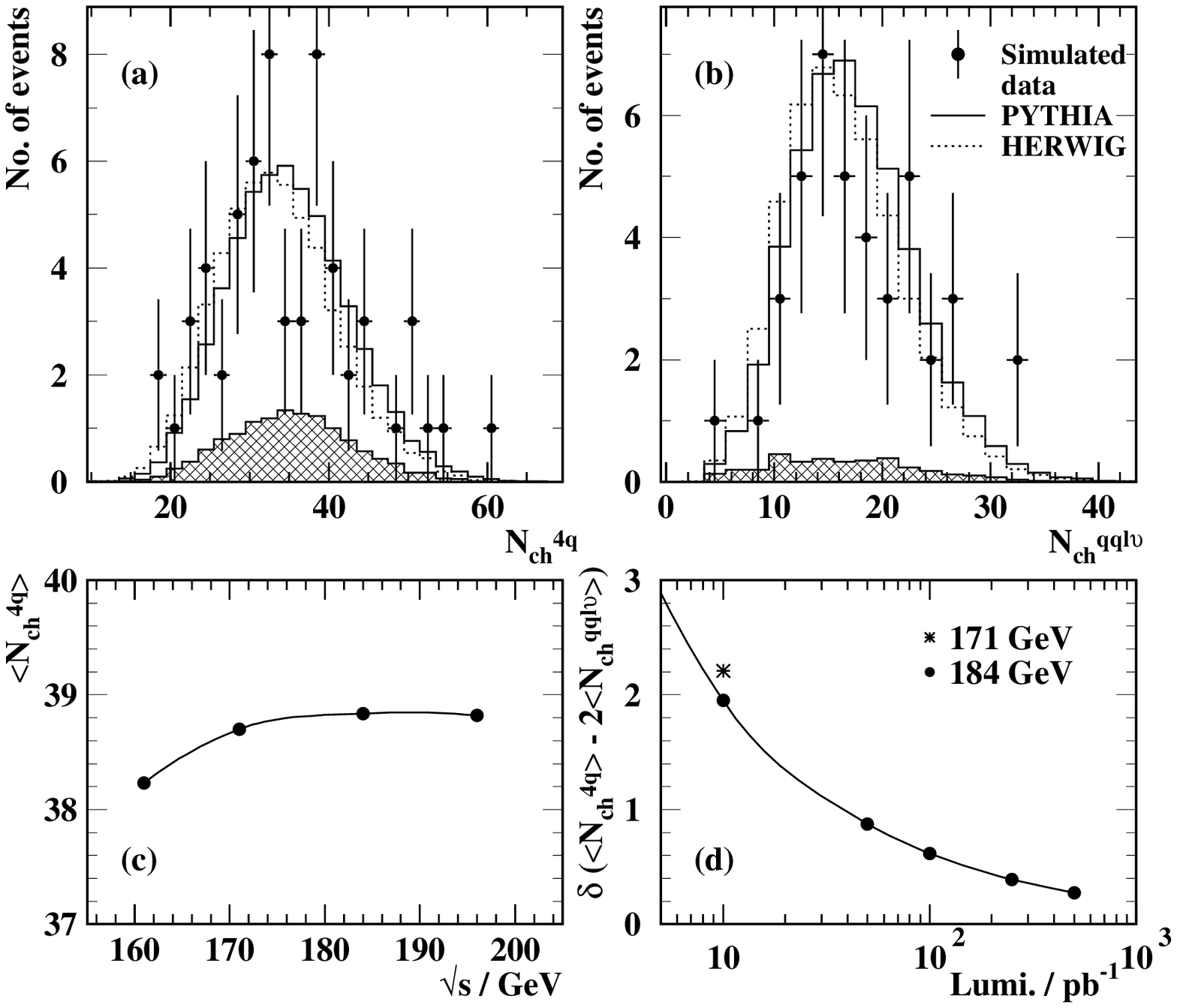}
 \caption{Inclusive charged multiplicity distributions with 10~pb$^{-1}$
   of fully simulated data, with background indicated hatched, at
   $\protect\sqrt{s}=171$~GeV for (a) \protect\WWqqqq, and (b)
   \protect\WWqqlv\ events.  (c) Variation of
   \protect\mnchqqqq\ with $\protect\sqrt{s}$.  (d)
   Luminosity dependence of the statistical uncertainty of
   $\protect\mnchqqqq-2\mnchqqlv$ (units of
   multiplicity).}
  \label{fig:cr_nch}
\end{figure}

The resulting (uncorrected) charged multiplicity distributions for the
hadronic and semi-leptonic channels are shown in Figs.~\ref{fig:cr_nch}(a)
and~\ref{fig:cr_nch}(b), respectively. The simulated data correspond
to an integrated luminosity of 10~pb$^{-1}$ at $\sqrt{s}=171$~GeV,
i.e. that delivered by \LEP\ in 1996. In both distributions, the expected
background is shown as a hatched histogram. The significant level of \Zqq\
background is apparent in the fully hadronic channel.

 To extract the mean charged multiplicity at the hadron level at a fixed
 centre-of-mass energy from such distributions, one can apply a simple
 correction, based on Monte Carlo, to the observed mean value, 
 after subtracting the expected
 background contribution. An alternative is to carry out a matrix-based
 unfolding procedure using the event-by-event correlation between the
 charged multiplicity at the hadron level and that observed in the detector
 after all analysis cuts have been performed. A separate correction for the
 effects of initial state radiation are necessary in this latter case.  A
 third alternative is to integrate the fragmentation function but this is
 not discussed here.
 
 Based on the the simulated data in Fig.~\ref{fig:cr_nch}(a) and (b), the
 expected statistical uncertainty on the difference
 $\mnchqqqq-2\mnchqqlv$ for an integrated
 luminosity of 10~pb$^{-1}$ is 2.2 units, or 5.7\% on
 \mnchqqqq. The evolution of the precision of such difference
 measurements with more data is estimated using the following assumptions.
 Firstly, the distributions of \nchqqqq\ and \nchqqlv\ are seen to be
 relatively insensitive to changes in centre-of-mass energy once away from
 the threshold region, as illustrated by the energy dependence of
 \mnchqqqq\ in Fig.~\ref{fig:cr_nch}(c). Therefore
 both the mean and the corresponding rms
 are assumed constant at their 184~GeV
 values. Secondly, above $\sqrt{s}=184$~GeV the \WW\ production
 cross-section is predicted to vary by less than 10\% in the region up to
 $\sqrt{s}<200$~GeV, and so a constant cross-section of 16~pb is assumed.
 Thirdly, it is assumed that the selection efficiency at 171~GeV may be
 maintained at higher energies. The expected background cross-section is
 not important as it is subtracted in performing the measurement.  Given
 these assumptions, the dependence of the expected statistical error on the
 difference, $\delta(\mnchqqqq-2\mnchqqlv)$, is
 shown as a function of integrated luminosity in Fig.~\ref{fig:cr_nch}(d).

 Typically in such multiplicity determinations, systematic effects become
 significant below a statistical precision of 0.5 units of multiplicity.
 Uncertainty in the modelling of 4-jet like \Zqq\ background with parton
 shower Monte Carlos in the
 fully hadronic channel may become a significant systematic.

\subsubsection{Event shapes}
Global event shape variables have been considered in earlier studies as
potential signatures for reconnection \cite{SK,GH,EG}. In most studies the
predicted effects on such observables induced by reconnection has been
sufficiently small that detection would be marginal,
even with an integrated luminosity of
500~pb$^{-1}$. 

The choice of a `no
reconnection' reference sample with which to compare data deserves some
thought.  In trying to find sensitive observables, using the models alone
is ideal. However, once possible signatures have been developed,
and one starts to search for effects in data, 
it will be invaluable
to have a well defined `no reconnection' reference sample in data 
to reduce model and tuning dependence.
\LEPONE\  data provide a high statistics reference, but additional assumptions
are necessary in either extrapolating energy scales, or in combining pairs
of \Zqq\ to emulate \WWqqqq\ events without reconnection.
It is also necessary to
assume that data recorded and processed by the detectors before 1996
can be directly compared with those recorded near the end of the \LEPTWO\
programme.  For some signatures, the ideal reference data are \WWqqlv\ 
events. However, this sample has only limited size and the comparison may
require the association of pairs of jets with Ws in the fully hadronic
channel, a procedure which experimentally introduces more uncertainty.  In
the following, all changes are relative to the `no reconnection' version of
each Monte Carlo model and all samples are \WWqqqq.

\begin{figure}[tb]
  \epsfxsize=\textwidth
  \epsffile{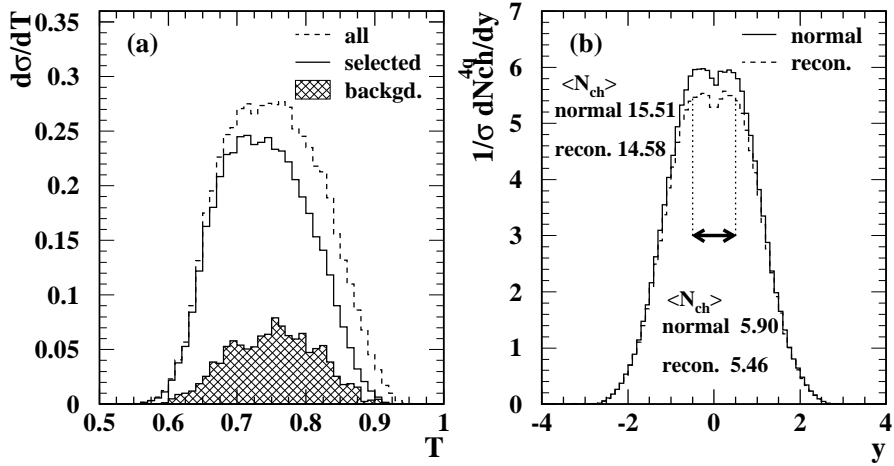}
 \caption{(a) Effect of typical experimental selection on thrust
          distribution, and (b) hadron level rapidity
          distribution in \ARIADNE\ for $p<1$~GeV.}
 \label{fig:cr_thrust_dndy}
\end{figure}

This study compares the differences in the rapidity distribution of charged
particles, \dndy, relative to the thrust axis of each event, in the central
region, $|y|<0.5$ and for all $y$, as suggested in \cite{GPZ,SK,GH}.  As
the effects are expected to be more pronounced for softer particles, the
distribution is studied for three momentum ranges, $p<0.5$~GeV, $p<1$~GeV
and all momenta.  It has been suggested \cite{GH,EG} that reconnection
effects may be more pronounced in specific topologies where the quarks from
different Ws are close to one another, therefore events are also studied
for all thrust values and for $T>0.76$.  One aspect not considered in
previous studies has been the effect of applying a realistic event
selection, which is necessary to reduce the large background ($\sigma(\Zqq)
\sim 20\sigma(\WWqqqq)$). As this is dominated by two-jet like events, the
efficiency for selecting \WWqqqq\ events in a similar configuration is
relatively small, as illustrated in Fig.~\ref{fig:cr_thrust_dndy}(a);
  $\sim38$\% of \WWqqqq\ events selected satisfy $T>0.76$, falling to
  $\lapproxeq 0.05$\% for $T>0.92$.

In \cite{EG}, the rapidity was studied relative to the axis bisecting the
two di-jet axes, as a function of the angle separating these axes.
Experimentally, without any reliable charge identification algorithm to
separate quarks from anti-quarks, the specific angle proposed in \cite{EG}
must at best be folded in experimental analyses, and also requires pairing
of jets into Ws. While the reliability of associating the `correct' jets
together is possible with moderate efficiency using kinematic fits,
selecting high thrust events was used in the current studies for expediency
and simplicity.  As the shifts in \MW\ expected are modest compared to the
experimental mass resolution on an event-by-event basis, it is worth
considering the use of kinematic fits in which our current knowledge of
\MW\ is applied as a constraint, in a similar way to that used by
experimental TGC analyses.

Hadronic events were generated using the models \PYTHIA, \HERWIG\ and \ARIADNE,
with and without a simulation of the \OPAL\ detector, and \dndy\ studied
within the ranges of $y$, $p$ and $T$ described above. A smearing
simulation of the \OPAL\ detector, which is reliable for studies in the
\WWqqqq\ channel and necessary to achieve the relatively high statistics
required, was used herein and also to estimate shifts in \MW.

\begin{figure}[tb]
%%% \centerline{\epsfig{file=dndy_trends.eps,width=\textwidth}}
 \epsfxsize=\textwidth
 \epsffile{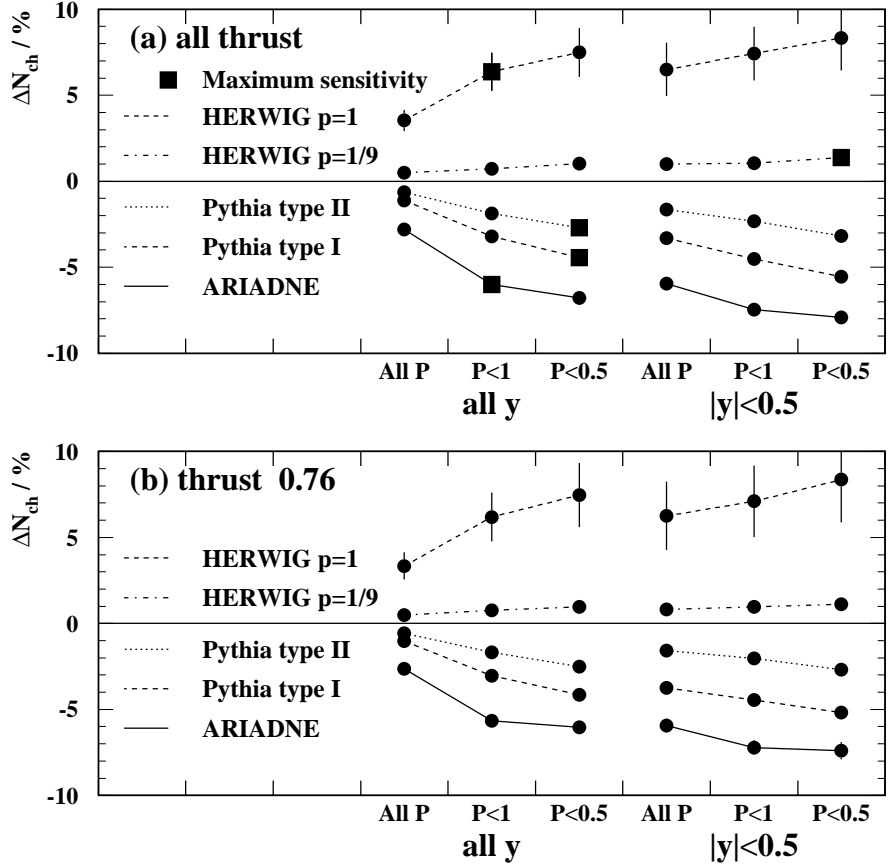}
   \caption{Fractional change in charged multiplicity as a function of
            maximum particle momentum, in two rapidity regions, for (a)
            all \protect{$T$}, and (b) \protect{$T>0.76$}. See text for
            details.}
   \label{fig:cr_dndy_trends}
\end{figure}

As an example of how the differences may be concentrated in restricted
rapidity intervals, Fig.~\ref{fig:cr_thrust_dndy}(b) shows the \dndy\ 
distribution for $p<1$~GeV in \ARIADNE, for events with and without
reconnection. Changes in charged multiplicity, \delnch, within given $p$
and $y$ intervals are summarised in Fig.~\ref{fig:cr_dndy_trends}(a) for
each of the models introduced in table~\ref{tab:cr_nch}, without detector
simulation.  The left (right) hand side of the figure shows the percentage
change in \mnchqqqq\ for the three momentum ranges considered
for all $y$ ($|y|<0.5$). The leftmost points in this figure correspond to
the results of table~\ref{tab:cr_nch}.  Fig.~\ref{fig:cr_dndy_trends}(b)
gives the analogous results for $T>0.76$.  For illustration, statistical
errors corresponding to an integrated luminosity of 500~pb$^{-1}$ are given
for the `\HERWIG\ colour octet' model.

It is seen that in all models the magnitude of the change increases when
only low momentum particles are considered. Applying a thrust cut such as
$T>0.76$ rejects $\sim40$\% of events and may change \mnchqqqq\ by up to
two units, but differences relative to the `no reconnection' scenarios are
essentially unchanged, therefore the sensitivity is reduced. The predicted
maximum statistical significance of \delnch, as well as its sign, depends
strongly on the model, varying from $\sim 6\sigma$ for \ARIADNE\ and the
\HERWIG\ `colour octet' model, $\sim 3.5\sigma$ for \PYTHIA\ type I, $\sim
2\sigma$ for \PYTHIA\ type II, down to $\sim0.8\sigma$ for \HERWIG. The
point of maximal sensitivity is indicated (square markers) for each model in
the figure. Similar trends were observed in studies with detector simulation
but typically \delnch\ was found to be $\sim$ 50\% smaller.

It may be possible to increase the sensitivity to reconnection effects
using charged multiplicity based methods, by considering particle
distributions relative to the \WW\ decay axis, as reconstructed using
kinematic fits. In \cite{YB}, an alternative multiplicity signature
(`interjet multiplicity') was introduced, having similar sensitivity to
integrating \dndy\ for $|y|<0.5$. This interjet multiplicity was similar in
idea to methods normally used to quantify the `string effect' in 3-jet
\epem\ events. It was suggested that this be studied further, using the
shape of the particle density distribution as a function of the angular
separation between jet pairs, rather than restricting the study to the
integrated particle density in the fixed angular regions. However, the
4-jet case is somewhat more complex than the familiar 3-jet case, being
non-planar, and so this was not pursued during the workshop.

\subsubsection{Shifts in \MW}

Extracting \MW\ from the decay products observed by experiments is
non-trivial, requiring much attention to bias induced from effects such as
initial state radiation, detector calibration, imperfect modelling of the
underlying physics processes and of the apparatus. In comparison to this,
estimating a shift which could result from the effects of reconnection
phenomena is straightforward, as the value of interest is the relative
shift between \MW\ determined in two different scenarios of the same
model. The absolute value of ``\MW'' obtained is not central to these
studies. However, there are still many uncertainties inherent in such
studies, such as sensitivity of the method used to extract \MW\ to changes
in $\sqrt{s}$, to tuning of the Monte Carlo models (e.g.  virtuality
cut-offs in the parton shower development), to treatment of combinatorial
background and ambiguous jet-jet combinations, and the range over which
fitting is performed to name but a few.

\begin{figure}[tb]
  \epsfxsize=\textwidth
  \epsffile{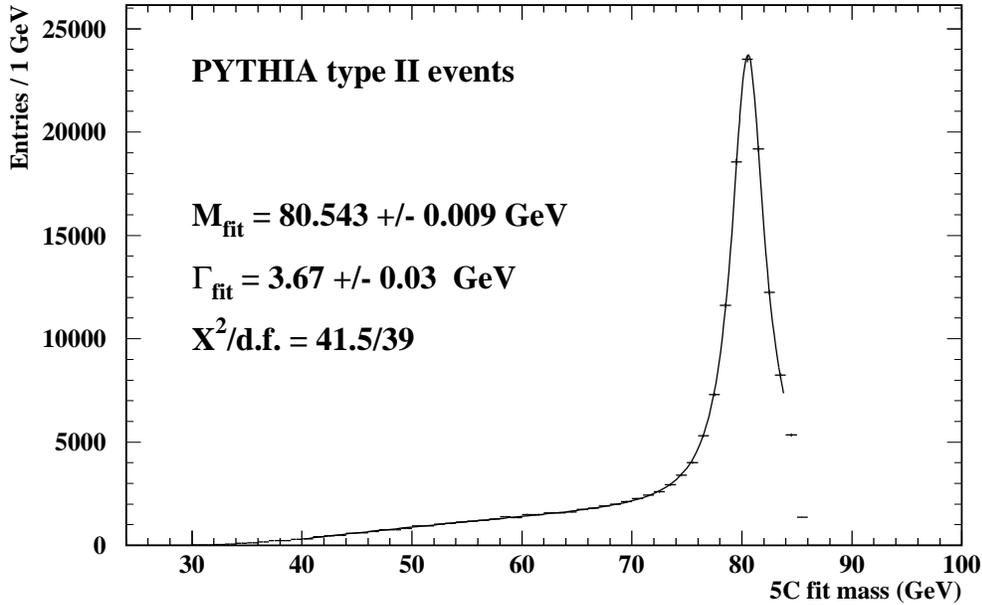}
  \caption{Typical mass distribution with fit results with
           detector simulation and event selection.}
  \label{fig:cr_mw_example}
\end{figure}

In these studies, the method used to extract \MW\ followed closely that
used by \OPAL\ for its preliminary \MW\ results using 172~GeV data. In this,
events with detector simulation are first selected using the same procedure
as noted earlier. Four jets are formed using the \kt\ jet finder, corrected
for double counting of energy within the apparatus, and a parametrisation
of the errors on the measured jet 4-momenta is carried out. A 
five-constraint kinematic fit, in which the jet-jet pair masses are constrained
to be equal, is performed for each of the three possible jet-jet pairings,
event by event. A mass distribution is constructed using
the mass from the combination having the highest probability from the
kinematic fit in each event if this has probability greater than 1\%. 
A second entry
is also admitted if the second most probable fit result has probability greater
than 1\% and within a factor of three of the highest probability combination.
The aim of this is to include additional mass information for events in
which the most probable fit combination is incorrect. In such events,
these two masses are essentially uncorrelated. A typical mass distribution
formed by this procedure is given in Fig.~\ref{fig:cr_mw_example}.

This method was applied to simulated events from each of the models in turn,
and the shifts obtained are summarised in table~\ref{tab:cr_deltamw}, where
uncertainties on these shifts are statistical.  The \ARIADNE\ model predicts
a modest shift in mass of approximately 50~MeV. No significant shift is
predicted by the models \PYTHIA\ and \HERWIG. In an earlier study, performed
in a similar way, significant shifts were determined \cite{YB}.  The
\PYTHIA\ and \ARIADNE\ models considered in the present study were also
included in \cite{YB}, albeit with different model dependent parameters and
looser event selection criteria.
\begin{table}
 \caption{Table of shifts in \MW\ for each model.}
 \label{tab:cr_deltamw}
 \begin{indented}
 \lineup
 \item[]\begin{tabular}{@{}llll}
 \br
       & & \multicolumn{2}{c}{$\protect\langle\Delta\MW\rangle$ (MeV)} \\
 model & & selected events ($\epsilon\simeq80$\%)  & all events \\
\mr
 \PYTHIA\  &  type I      &  $+18\pm$11 & $+11\pm$11 \\
         &  type II     &  $-13\pm$11 & $-19\pm$11 \\
 \HERWIG\  &  reconnected ($P=\frac{1}{9}$) & $-16\pm$16 & $-19\pm$16 \\
         &  reconnected ($P=1$)           & $+13\pm$15 & $+8\pm$14 \\
 \ARIADNE\ &  reconnected &  $+51\pm$16 & $+51\pm$15 \\
\br
\end{tabular}
\end{indented}
\end{table}

One quite plausible explanation proposed was that the difference was due to
the significantly more stringent event selection currently used. It has
been shown that the current selection preferentially rejects events having
two-jet like characteristics, which is where reconnection effects may be
expected to be prevalent. The rejection of these events does not appear to
be the reason for small mass shifts, as a similarly small effect is observed
when all events are selected, as seen in table~\ref{tab:cr_deltamw}.

Many possible sources for the difference were investigated in the context of
the \PYTHIA\ models. Neither changes in the tuning of \PYTHIA/\JETSET\ by
\OPAL\footnote{Among these, the cut-off parameter, $Q_0$, was increased
  from 1.0~GeV in the similar investigation of \cite{YB}, to 1.9~GeV.}  to
improve the description of \LEPONE\ data, nor the different centre-of-mass
energy ($\sqrt{s}=175$~GeV in \cite{YB}) were found to be significant.  The
current analysis procedure is slightly different to that in \cite{YB}.
However, significant shifts are still found when the current procedure is
applied to the same simulated events used in \cite{YB}.  Conversely,
applying the former procedure of \cite{YB} to the samples herein does not
induce a significantly larger mass shift.

One apparently significant effect was found to be the choice of mass
assigned to jets in performing kinematic fits. As discussed in \cite{YB},
this choice is not unique.  In the analysis of \cite{YB}, the hadronic jets
were assumed massless whereas in the current studies, the measured jet mass
was used.  Re-analysing the same simulated events of \cite{YB} but assigning
measured masses to the jets reduces the mass shifts estimated, e.g.\ shifts
quoted in \cite{YB} of $130\pm40$~MeV (type I) and $50\pm40$~MeV (type II)
become $70\pm40$~MeV and $30\pm40$~MeV, respectively. For comparison, a
sample of 200\,000 fully hadronic type I events were generated at
$\sqrt{s}=175$~GeV using {\em identical\/} model parameters and program
versions, and analysed using the procedure of \cite{YB}, also using measured
jet masses. This yielded an estimated shift of $46\pm16$~MeV.  It should be
noted that fluctuations due to finite Monte Carlo statistics have to be
considered when comparing with the results of \cite{YB}, in which samples
sizes for the analogous studies were 50\,000 events.

Comparing the results for mass shifts in table~\ref{tab:cr_deltamw}
with multiplicity shifts in table~\ref{tab:cr_nch}, it can be seen
that any relationship between them is model dependent. Furthermore,
relatively large shifts in the charged multiplicity do not necessarily
lead to a significant shift in \MW.

\subsubsection{Future}
The future for experimental studies of colour reconnection is quite open.
There is clear model dependence in signatures and mass shifts may be
smaller than earlier proposed \cite{YB}, although there are other
models available \cite{NOVA,EG} which were not tested in this study from
which different conclusions may be drawn. A necessary condition for a model
to be taken seriously is that it should describe the data, therefore
 tuning of models has to be addressed. With the current statistical
precision of \LEPTWO\ data, none of the models has been put to a stringent
test. The effect of background cannot be ignored in the \WWqqqq\ channel
as it proves difficult to remove. More sophisticated selections may
be developed, but typically these make use of non-trivial correlations
between observables, which may be poorly described by the models.
A particular concern is the description of parton shower Monte Carlos
to describe the hard, 4-jet like background which is selected.
The remaining point of note is that given the model dependence inherent
to such studies, it is most important to develop signatures which can
be tested taking the `no reconnection' scenario from data themselves.

\setcounter{footnote}{3}
\subsection{Bose-Einstein correlations\/\protect\footnotemark[\value{footnote}]}
\footnotetext[\value{footnote}]{Prepared by V~Kartvelishvili, D~R~Ward}
\label{bose-eins}

As discussed in \cite{BW}, studies of the influence of Bose-Einstein 
correlations on the W mass measurement were carried out for the 
\CERN\ \LEPTWO\ workshop~\cite{YB,tt10}, which
suggested that there could be sizeable shifts in the W mass,  
$\Delta M_{\mathrm{W}} \sim {\cal O}(+100\mev)$.
These studies were based on the \verb|LUBOEI| algorithm, which shifts 
particle momenta after generation of the event, and thus requires
rescaling.  In this report of the present workshop, 
we outline some more recent results using event
weighting schemes.

%prepared by V.Kartvelishvili for D.Ward - W mass working group report
%
\setlength{\unitlength}{1mm}
\subsubsection{Event weighting schemes for Bose-Einstein effects}

The Bose-Einstein effect corresponds to an enhancement
 in the production probability for identical bosons to
 be emitted with small relative momenta, as compared
 to non-identical particles under otherwise similar conditions.
 Assuming a spherical space-time distribution of the particle source,
 the correlation function takes the form:
\begin{equation}\label{c1}
 C(Q) = 1 + \lambda \rho (Q)
\end{equation}
where $Q$ is the four-momentum difference, $Q^2 = -(p_1-p_2)^2$,
 and $\rho$ is the absolute square of the Fourier transform 
 of the particle emitting source density, with the normalization condition
 $\rho(0)=1$.
 The incoherence parameter $\lambda$ takes
 into account the fact that, for various reasons, the strength of
 the correlations can be reduced. 

Often a Gaussian model is assumed for the source density, which leads to
\begin{equation}\label{Gauss}
 \rho (Q) = \exp(-R^2 Q^2)
\end{equation}
 where $R$ is the source radius. 

In {\cite{kkm}} the problem of
 Bose-Einstein correlations was addressed using an approach
 based on assigning weights to the simulated events
 according to the momentum distributions of final state bosons.
 In this global event weight scheme a shift in the reconstructed
 W mass distribution may arise
 if the average event weight depends on \MW.
 The use of global event weights is 
 complementary to the local reweighting scheme of ref.~\cite{tt10} in 
 the sense that as opposed to the latter, in the former the kinematical 
 properties of the events are preserved, while all probabilities 
 and multiplicities may change. 
 The method arises very naturally in a quantum mechanical approach, 
 where the weight can be constructed as the ratio of the square of the 
 symmetrized multiparticle amplitude to the square of the 
 non-symmetrized amplitude
 corresponding to the emission of distinguishable particles. 
 However, the use of global event weights leads
 to a number of conceptual and computational difficulties, which
 must be overcome before any quantitative conclusions can be drawn.
 
One way of forming the weight is to take a product of
 enhancements $C(Q)$ for all pairs of identical bosons
 in the event:
\begin{equation}\label{v1}
 V_1 = \prod_{{i_1,i_2}}C({Q_{i_{1}i_{2}}})
\end{equation}
For high multiplicity events this weight can become extremely large,
 so that a few such events dominate the weighted distributions and
 lead to unrealistic results. The event weights therefore
 have to be regularized in some way.
 In order to keep the statistical error at a reasonably
 low level, events with very high weights (higher
 than some $V_{\mathrm{max}}$) were discarded. The resulting dependence upon
 $V_{\mathrm{max}}$ 
 was analysed and the results were extrapolated to $V_{\mathrm{max}} \to \infty$.

One can also rescale the weight of the event using a single
 constant $W_0$:
\begin{equation}\label{v2}
 V_2 = V_1/W_0^n
\end{equation}
 where $W_0$ is slightly larger than 1, and $n$ is the
 number of pairs in the event (i.e. the number of terms
 in the product in (\ref{v1})). 
 However, for $V_{\mathrm{max}}$ fixed and reasonable values of $W_0$
 the scheme gave rise to numerical difficulties, stemming
 from the fact that increasing $W_0$ brings in more and more events from
 the high weight tail of $V_1$, which leads to large fluctuations.
 The results for the shifts in multiplicity and \MW\ using $V_2$ were
 roughly consistent with those found for $V_1$.
\begin{figure}[tb]
  \begin{center}
    \begin{picture}(100,100)(0,0)
    \epsfig{file=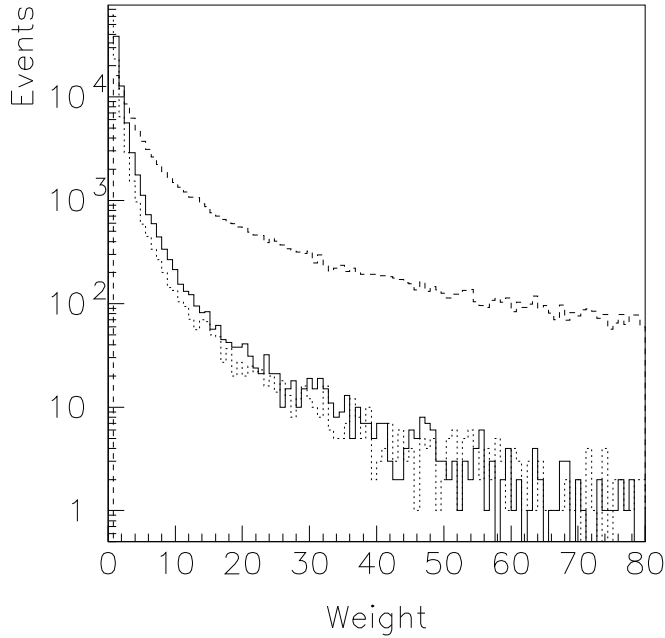,height=100mm}
    \end{picture}
  \end{center}
\caption{Distributions of the weights $V_1$ (dashed line), $V_3$ 
(full line) and $V_5$ (dotted line) for \WW\ events at 175 GeV.}
\label{fig-vato1}
\end{figure}

A general problem with any weighting method is, that like the local 
 reweighting scheme of ref {\cite{tt10}}, they introduce
 artificial correlations also between non-identical particles.
 In order to counteract this, one can rescale $V_1$ using
the weight calculated with 
 non-identical pairs:
\begin{equation}\label{v3}
 V_3 = V_1/V_0^{n/m}
\end{equation}
where $V_0$ is the weight calculated according to (\ref{v1}) but for
 non-identical bosons in the same event, while $n$ and $m$ are
 the numbers of identical and non-identical pairs, respectively.
 This also leads to a better numerical behaviour, 
 as illustrated in Fig.~\ref{fig-vato1}, which shows
 the distributions of $V_1$ and $V_3$ for simulated \WW\ events at 
 $\sqrt{s} = 175$ GeV. The high weight tail is much less
 pronounced for $V_3$ than for $V_1$.
 Both fall off to a good approximation as inverse powers,
 with exponents $-2.6$ and $-1.4$, respectively, which makes it plausible
 that the sum of all weights converges.

A different method of constructing the event weight, which is closer to
 a full quantum mechanical treatment, starts from the
 introduction of a symmetric amplitude, which has $n!$
 terms {\cite{tt14}}. This leads to a weight:
\begin{equation}\label{v4}
 V_4 = \sum_{{{\rm{permutations}}}} \lambda^{k/2}{\rho}(Q_{1i_1})
     {\rho}(Q_{2i_2}) \cdots {\rho}(Q_{ni_n}),
\end{equation}
where $k$ is the number of times when the first and
 second indices differ. For $Q=0$, $\lambda=1$
 and $n$ identical particles, equ. (\ref{v1}) gives
 a weight of $2^{n(n-1)/2}$,
 while equ. (\ref{v4}) results in the correct  value $n!$.
 However, for typical hadronic configurations this difference
 is much smaller, and (\ref{v4}) is rarely used because of
 computational difficulties: by limiting the number of identical
 particles of each type to 8, one loses about 30\% of events at
 the Z peak and about 50\% of events in W pair production.

In a recent attempt to overcome these difficulties, in {\cite{fial}},
 the $n!$ permutations in (\ref{v4}) were divided into sub-classes where 
 exactly $k$ particles have been exchanged, and for low
 energy hadronic collisions ($\sqrt{s}=10-30$ GeV) 
 BE effects were shown to saturate already at $k=5$.
 However, possible reasons for this improvement are that the 
 multiplicities in hadronic collisions in the above energy range are
 much lower than at \LEP, and that the source
 radius used in {\cite{fial}} was $R=1$ fm as opposed to 0.5 fm measured
 at the Z peak at \LEPONE\ and used in {\cite{kkm}}. The larger radius 
 corresponds to
 a narrower distribution in $Q$\/-space, so that the peak in
 the $Q$\/-distribution of pions does not contribute any longer
 and the number of significantly contributing pairs is dramatically reduced.
 A subroutine \verb|LWBOEI| for calculating event weights using this approach
 can be incorporated into \JETSET/\PYTHIA, but it is not clear yet
 how well this approach works for \ee\ collisions at \LEP\ energies.

Similarly, another simplified version of (\ref{v4}) was studied in
 {\cite{jada}} to assess BE effects on W mass measurements at \LEP.
 Here, particles were divided into ``clusters'' of neighbours in $Q$\/-space,
 and simple formulae were derived from (\ref{v4}) under certain
 assumptions (see {\cite{biaj}} for details). 
 Detector effects and reconstruction procedures have also been included.
 The authors of {\cite{jada}} also use the source radius of 1 fm,
 and do not see any W mass
 shift due to BE correlations at the level of their statistical precision,
 concluding that BE effect has a negligible influence, below 30 MeV,
 on the reconstruction of W mass.

All event weights considered above were based on the
 Gaussian parametrization (\ref{Gauss}) of the particle emitting source.
 This implies that $V_1 \geq 1$ for all values of $Q$.
 In addition to the above weights, a different pair weight was also
 studied in {\cite{kkm}}, inspired by recent theoretical 
 studies \cite{bo}.
 Here $\rho$ in (\ref{c1})
 is not required to be always positive: 
\begin{equation}\label{wb}
 \rho(Q) = {{\cos (\xi Q R)}\over{\cosh (Q R)}}
\end{equation}
For $\xi$ close to 1, this is very close to (\ref{Gauss}) apart from
 becoming slightly negative at large $Q$. The corresponding weight
 $V_5$ was built in analogy with (\ref{v1}), but with the Gaussian 
 (\ref{Gauss}) replaced by (\ref{wb}) (the dotted line in Fig.~\ref{fig-vato1}). 
 $\xi=1.15$ was found to 
 lead to a good overall description. Due to the better
 numerical behaviour of this weight function (the exponent in the
 power fit is $-2.4$), it was possible to apply $V_5$
 without further rescaling.

\subsubsection{Influence of event weighting on $\mathrm{Z}$ properties}

 In order to
 check the self-consistency and inherent systematic errors of the method 
 the same
 weighting procedure was applied to the well studied process of Z 
 hadronic decays.
 For weight calculation, $\lambda =1$ was used for pions and kaons 
 originating from sources
 with decay lengths $c\tau < 10$~fm, and $\lambda = 0$ otherwise.
 The source radius $R$ was taken to be equal to 0.5~fm everywhere.
 In Z decays at \LEPONE\ one observes $\lambda \approx 0.3$ if all
 particles are considered, 0.4 if only pions are taken into account
 and 1.0 for directly produced pions, while $R \approx 0.5$~fm
 {\cite{alephBE,delphiBE}}.

Various measurable properties of the Z will be affected
 to different extents, if one introduces event weights into the
 simulation of its hadronic decays. Since the partonic states 
 before hadronization are known to be well described by
 perturbative calculations, which do not take into account
 Bose-Einstein correlations, uncritical application of
 event weights may lead to large inconsistencies with e.g.\
 measured branching ratios and relative frequencies of jet 
 multiplicities etc. In order to see how serious these effects
 are and to judge what consequences this has for the
 analysis of the \WW\ events, the precise experimental
 data from Z decays can be used to check the event
 weighting schemes of Bose-Einstein correlations.
 Samples of 100000
 hadronic events at $\sqrt{s}=M_{\mathrm{Z}}$ and $M_{\mathrm{Z}} \pm 2$~GeV
 were simulated and the weighting schemes described above were applied.
 Table~\ref{tab-vato1} presents the changes in the charged particle multiplicity,
 in the apparent Z peak position in hadronic vs leptonic decay modes,
 in the branching fractions for charm and beauty decays
 ($R_c$ and $R_b$) and in the ratio of three- to two-jet events 
 with and without event weighting.

\begin{table}
\caption{
 Differences in charged multiplicity, apparent peak mass of Z, branching
 fractions and three-to-two jet event ratio, between weighted
 and non-weighted events, for various weighting systems described
 in the text. 
}
\label{tab-vato1}
\lineup 
\smallskip
\begin{indented} 
\item[]\begin{tabular}[tbc]{lrrr} 
\br
        & $V_1$  & $V_3$  & $V_5$  \\
\mr
$\Delta \langle n_{ch} \rangle$   & $3.7 \pm 0.5$
       & $1.3 \pm 0.2$ & $1.8 \pm 0.2$ \\

$\Delta M_{Z_0}$,~MeV & $8\pm 3$ & $  0\pm 3$ &      $ 1\pm 4$ \\

$\Delta R_c,$ \%      & $-3\pm 2$ & $  -2\pm 2$ &      $0\pm 2$ \\

$\Delta R_b,$ \%      & $-26\pm 3$ & $-11\pm 2$ &      $-5\pm 2$ \\

$\Delta$ 3jet/2jet, \%      & $80\pm 20$ & $  20\pm 5$ &  $ 20\pm 5$ \\
\br
\end{tabular}
\end{indented}
\end{table}

This analysis resulted in the following: 
\begin{itemize}
\item
 The average charged multiplicity has changed. The weight
 $V_1$, which was not rescaled, leads to the largest
 increase when compared to the unweighted results, while both $V_3$ 
 and $V_5$ give a smaller increase around 1.5. In all these cases, 
 the change can be accommodated by
 retuning the parameters in the simulating program.
\item
 In principle, event weighting can result in a shift of 
 the apparent Z mass peak.
 However, only $V_1$ yielded a shift of a few MeV, while for 
 $V_3$ and $V_5$ the shift is essentially zero.
 We have not found any significant change of the apparent Z width.
\item
 The pattern of heavy and light quark fragmentation is rather different.
 Heavy quarks produce significantly fewer pairs with small $Q$, and all BE
 effects in this approach are less pronounced for heavy
 quarks. Heavy quark events thus obtain smaller average weights, which
 result in changes shown in Table~\ref{tab-vato1}. Note that the effect for $c$-quarks
 is diluted because $b$-quark events reduce the overall average 
 weight. In order to exclude this artificial 
 flavour dependence in W decays, the weighting
 and rescaling was performed separately for the different decay modes of the
 Ws.
\item
 The weighting resulted in a
 substantial increase of jet activity, as measured by the three to
 two jet event ratio. This is however difficult to
 quantify because of its dependence upon the jet finding algorithm and its
 parameters. The numbers shown in Table~\ref{tab-vato1} were obtained using
 \verb|LUCLUS| with default parameters ($d_{\mathrm{join}}=2.5$~GeV), 
 corresponding to fairly 
 narrow jets. The effect decreases for broader jets and in any case is 
 much less pronounced in \WW\ production, so no attempt was made to
 correct for it.
\end{itemize}

\begin{figure}[tb]
  \begin{center}
    \begin{picture}(140,100)(0,0)
%      \put(0,100){\special{insert size 140 100
%            \the\unitlength\space
%               cq.eps}}
    \epsfig{file=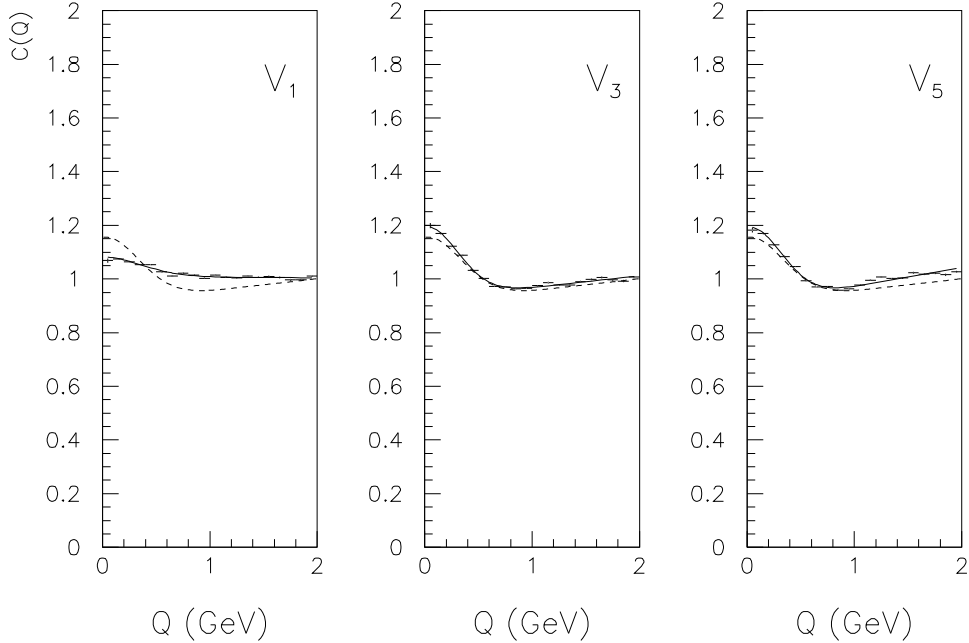,height=100mm}
    \end{picture}
  \end{center}
\caption{Reproduced correlation functions in Z events using the 
weights $V_1$, $V_3$ and $V_5$. The dashed lines show the result of a 
fit to real data from hadronic Z decays
\protect\cite{delphiBE}.}
\label{fig-vato2}
\end{figure}

The reproduced correlation functions for the three weighting schemes, 
 $V_1$, $V_3$ and $V_5$ are shown in Fig.~\ref{fig-vato2}.  
 Also shown are fits to the form
 \begin{equation}
   N(1 + \beta Q)(1 + \lambda \exp (-Q^{2}R^{2}))
 \end{equation}
 which is often used to parametrize the experimentally observed correlation
 function in Z decays \cite{alephBE,delphiBE}. The dashed
 line in each figure represents the result of a fit to the correlation function
 of all particles observed in real data from hadronic 
 Z decays {\cite{delphiBE}}.

$V_1$ does not reproduce the correlation function well, giving
 too small values for both the incoherence parameter $\lambda$ and the
 radius $R$, while $V_3$ and $V_5$ both give very reasonable 
 descriptions.

Hence one concludes that, provided that the different quark final states
 (and possibly the final states with different number of jets)
 are treated separately, application of the global event weighting
 technique with rescaling of the weight $V_3$, or $V_5$ built using 
 the pair weight ({\ref{wb}})
 is not inconsistent with \LEPONE\ data at the Z, whereas the direct
 application of the product of Gaussian pair weights $V_1$ should be 
 treated with more care.

\subsubsection{W pair production}

\PYTHIA\ 5.7~\cite{tt11} was used to simulate the 
 process $\ee\rightarrow\WWqqqq$,
 and the weighting schemes described above were applied to simulate BE
 correlations.
 A basic assumption here is that hadronic W and Z boson decays
 are sufficiently similar, so that by using the tuning of the
 Monte-Carlo model parameters that reproduces the experimental
 data from Z decays at \LEP, Bose-Einstein effects in single
 W decays are already effectively taken into account in
 properties such as multiplicities and single particle momentum
 spectra.
 Only correlations between identical bosons originating
 from different Ws were included, since BE correlations
 within a single W cannot lead to any 
 change in the W mass compared to the semileptonic
 channel  $\ee\to\WWqqlv$. 
 Measurement effects like experimental resolution
 and acceptance, reconstruction method etc. were not taken into
 account.

The information from the Monte-Carlo was
 used to assign each final particle to the W$^+$ or the W$^-$,
 as in {\cite{tt10}}.
 Ws with mass values in the interval
 $70~{{\rm GeV}} \leq M_{W} \leq 90$~GeV
 were  studied at 175 and 192~GeV to assess the
 energy dependence. At each energy, $10^5$ events were generated,
 which is about an order of magnitude higher than the
 expected statistics of all four \LEP\ experiments combined at
 500~pb$^{-1}$ integrated luminosity per experiment.
 In general, one expects that BE-induced effects in \WW\ production should
 die out at high energies, as the overlap between the two W decay
 volumes decreases. This requires much higher energies than will become
 available at \LEPTWO, however, and it is likely that
 the effect will increase with energy in the \LEPTWO\ range \cite{tt10}.

The mass distribution of W bosons was built with and
 without event weighting for each of the weights used,
 and the differences were calculated
 in the average charged multiplicity $n_{ch}$,
 the mean W mass, $M^{\rm{mean}}_W$,
 averaged over the whole interval $70~{{\rm GeV}} \leq M_{W} \leq 90$~GeV,
 and a fitted $M^{\rm{fit}}_W$.
 The fit was performed using a relativistic Breit-Wigner shape with an
 $s$-dependent width, in the interval $ 80.25\pm \delta$~GeV, with
 $\delta = 2$ GeV. The results are presented in Table~\ref{tab-vato2}.

\begin{figure}[tb]
  \begin{center}
    \begin{picture}(140,100)(0,0)
    \epsfig{file=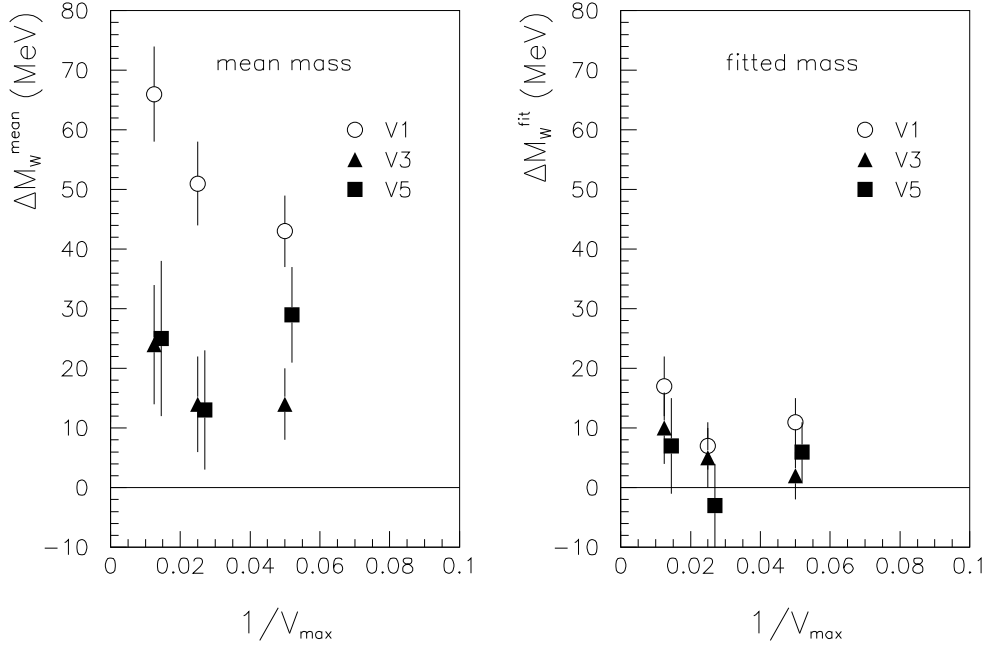,height=100mm}
    \end{picture}
  \end{center}
\caption{
Shifts in the mean and fitted W mass as functions of
the inverse weight cut off, $1/V_{\mathrm{max}}$, for weight schemes $V_1$, $V_3$ 
and $V_5$, and $\protect\sqrt{s} = 175$ GeV.
}
\label{fig-vato3}
\end{figure}

As mentioned above, for computational reasons, 
 events with very large weights have been discarded.
 The dependence on the cutoff value, $V_{\mathrm{max}}$, was eliminated by 
 calculating the multiplicity and mass shifts for
 three values of $V_{\mathrm{max}}$ (20, 40 and 80) and then extrapolating to
 infinite cutoff. This method seems to be more reliable and less
 vulnerable to fluctuations than direct calculation with very high
 $V_{\mathrm{max}}$.
 Fig.~\ref{fig-vato3} shows the values of the mean W mass, $M^{\rm{mean}}_W$,
 and the fitted $M^{\rm{fit}}_W$ as functions of $1/V_{\mathrm{max}}$. 
 For $V_1$, the extrapolated value depended
 on the specific way the extrapolation was performed, and this
 ambiguity was added to the error shown in Table 4.

\begin{table}
\caption{
Values of differences in multiplicity and mass of the W boson for events
 with and without interconnecting Bose-Einstein correlations between the
 two Ws. 
}
\label{tab-vato2}
\lineup 
\smallskip
\begin{indented} 
\item[]\begin{tabular}[tbc]{lrrr}
\br
                & $V_1$  & $V_3$  & $V_5$  \\
\mr
$\Delta n_{ch}$    &          &          &        \\

 175 GeV           &  $3.8 \pm 0.5$ &  $1.8 \pm 0.2$  & $ 1.0 \pm 0.2$ \\
 192 GeV           &  $3.7 \pm 0.5$ &  $1.7 \pm 0.2$  & $ 0.6 \pm 0.2$ \\
\mr

$\Delta M_W^{mean}$~(MeV)  &     &          &        \\

  175 GeV          &  $75 \pm 15$  &  $22 \pm 11$  &  $20 \pm 14$ \\
  192 GeV          &  $92 \pm 16$  &  $34 \pm 11$  &  $38 \pm 14$ \\
\mr

$\Delta M_W^{fit}$~(MeV) &       &          &       \\

 175 GeV          &  $12 \pm  9$  &  $11\pm 7$    &  $ 4 \pm 12$ \\
 192 GeV          &  $15 \pm  8$  &  $13\pm 7$    &   $6 \pm  9$ \\
\br
\end{tabular}
\end{indented}
\end{table}

From these numbers one can draw the following conclusions:

\begin{itemize}

\item
There is a clear correlation between the BE-induced shifts in the W mass
 and in the charged particle multiplicity in \WW\ production: the larger
 the increase in charged multiplicity, the larger are the expected mass shifts. 

\item
Both $V_3$ and $V_5$ weights result in fairly
 small mass shifts.
 They are well-behaved numerically 
 and probably give quite reliable estimates of the effect.
 The spread of values using various weighting schemes can be
 considered as indicative of
 the systematic errors inherent to this approach. 

\item
The fitted value for the W mass is less sensitive to BE effects
 than the mean over the full distribution, which has been used
 to estimate the effect in previous investigations \cite{tt10}.
 The estimated values for the shift in the fitted mass are
 less than 20 MeV, implying that BE correlations are not too
 dangerous for the W mass measurements at the expected level of accuracy
 at \LEPTWO. For the shifts in the mean W mass, values of order of a few
 tens of MeV were found, of the same general magnitude as in {\cite{tt10}}.
 In all cases the shift is 
 towards larger masses, as expected on general grounds \cite{YB,tt10}.

\item
For all weighting schemes, the shift in \MW\ increases
 with energy in the energy range considered, but the increase is fairly
 small.

\end{itemize}

It is interesting to compare these results to the
 predictions based on the implementation of Bose-Einstein
 effects by shifting the momenta of final state particles
 {\cite{tt10}}. 
 The most important difference is in the particle multiplicity:
 event weighting naturally leads to an increase of the average number of
 particles due to Bose-Einstein correlations, while the
 momentum-shifting method assumes that the multiplicity is
 unchanged. 
 The energy dependence of W mass shift is also different.
 The strong energy dependence in  momentum-shifting scheme
 is a combination of two effects: the increase of the systematic
 shift for low momentum particles in the direction of smaller
 W momenta, and the differences in momentum spectra of
 W decay products for various energies, as stressed
 in {\cite{tt10}}. This seems to be less pronounced in the present
 approach.

The study {\cite{kkm}} confirms that the systematic effect of BE correlations
 on the W mass determination can potentially be quite large, as found
 in \cite{tt10}, although the actual values of the
 mass shift found here are somewhat smaller. The size of the shift is
 however quite sensitive to the procedure used to extract the value of the
 W mass. In particular it is observed that a fit to the
 lineshape of the W mass distribution has a much smaller systematic error 
 from Bose-Einstein correlations than the average mass, due to the
 fact that the main effect on \MW\ in our scheme arises from the tails
 of the mass distribution, which contain very little information about
 the peak position. Hence it seems possible to keep the systematic 
 error from this source below about 20 MeV. Careful work linked 
 to the actual event selection and fitting procedures used
 by the \LEP\ experiments is obviously needed in order to assess this in 
 the individual cases and to optimize the analysis procedures.
 Since the value of the mass shift
 is always positive (as also expected on general grounds), a further
 reduction of the systematic error by a factor two is in principle
 possible by assigning the expected shift as a correction to \MW.

The comparison between hadronic decays in the
 \WWqqqq\ and \WWqqlv\  
 channels gives a unique possibility to investigate the
 influence of Bose-Einstein correlations on various properties
 of final state particles, such as multiplicity, transverse
 and longitudinal momentum spectra, resonance properties and reconstructed
 jet characteristics. It is possible, that by taking proper care
 in the fitting procedures used, one may at the
 same time be able to use a large part of the hadronic \WW\ events for the
 W mass determination, and to study the interconnection
 effects in the relatively clean setting of \ee\ $\to$ \WW\ events, 
 by restricting the study to the region of large, off-peak W masses 
 where these effects are expected to be the largest.

\newpage

%%%%%%%%%%%%%%%%%%%%%%%%%%%%%%%%%%%%%%%%%%%%%%%%%%%%%%%%%%%%%%%%%%%%%%%%%%
%               the FOUR-JET subsection %%%%%%%%%%%%%%%%%%%%%%%%%%%%%%%%%%%%%%%%%%%%%%%%%%%%%%%%%%%%%%%%%%%%%%%%%%

\setcounter{footnote}{3}
\subsection{Four-jet events\/\protect\footnotemark[\value{footnote}]}
\footnotetext[\value{footnote}]{Prepared by R Jones,  S Moretti and 
W J Stirling}
\label{subsub3}

\newcommand{\qqgg}{\mbox{$\mathrm{q\overline{q}gg}$}}
\newcommand{\MC}{\mbox{\sc MC}}
\newcommand{\PaSh}{\mbox{\sc PS}}

The \QCD\ processes $\epem \to (\Zz/\gamma)^* \to \qq\qq, \qqgg$ form significant
backgrounds ($\gapproxeq 20\%$)  to 
$\WW \to 4\;$jet production at \LEPTWO. 
It is important, both 
for the threshold and direct reconstruction \MW\ measurements, that these
backgrounds are well under control. In particular, the \QCD\ Monte Carlos
(\MC s) used in the \MW\ analyses should correctly describe the relevant features
(for example, the overall rate and the kinematic distributions) 
of the four-jet final states which pass the \WW\ selection criteria.

In this connection it is worrying that certain aspects of four-jet production
are {\em not} well described by the standard `parton-shower (\PaSh) $+ O(\alpha_s)$'
 \MC s 
(\JETSET, \HERWIG, \ldots).
As discussed by G~Cowan at this Workshop (\cite{GC}, in particular
Fig.~9,  see also
\cite{ALEPHgluino}), four-jet studies
performed by the \ALEPH\ collaboration at \LEPONE\ reveal significant disagreement
between data and \MC s for distributions in the standard four-jet shape variables  
$\chi_{\mathrm{BZ}}$, $\phi_{\mathrm{KSW}}$, $\theta^*_{\mathrm{NR}}$ and  $\alpha_{34}$ (for definitions, see for example Ref.~\cite{ALEPHgluino}).
This suggests that the \MC s do not  provide a correct description of
the angular correlations between the quark and gluon jets.
On the other hand, ${\cal O}(\alpha_s^2)$ matrix element models
(e.g. the \JETSET\ ${\cal O}(\alpha_s^2) + \mbox{string fragmentation}$ model)
 give a much
better description of four-jet final states \cite{GC}. The problem here is  that  
matrix element models with `added-on' hadronisation cannot be reliably extrapolated
from \LEPONE\ to \LEPTWO\ energies -- the hadronisation tuning is only valid at
the lower energy. (It was reported at the Workshop that \ALEPH\ have a special
`${\cal O}(\alpha_s^2) + \PaSh + \mbox{cluster hadronisation}$' 
 version of \HERWIG\ , but this is so far not publicly available.)

\begin{figure}[tb]
  \epsfxsize=\textwidth
  \epsffile{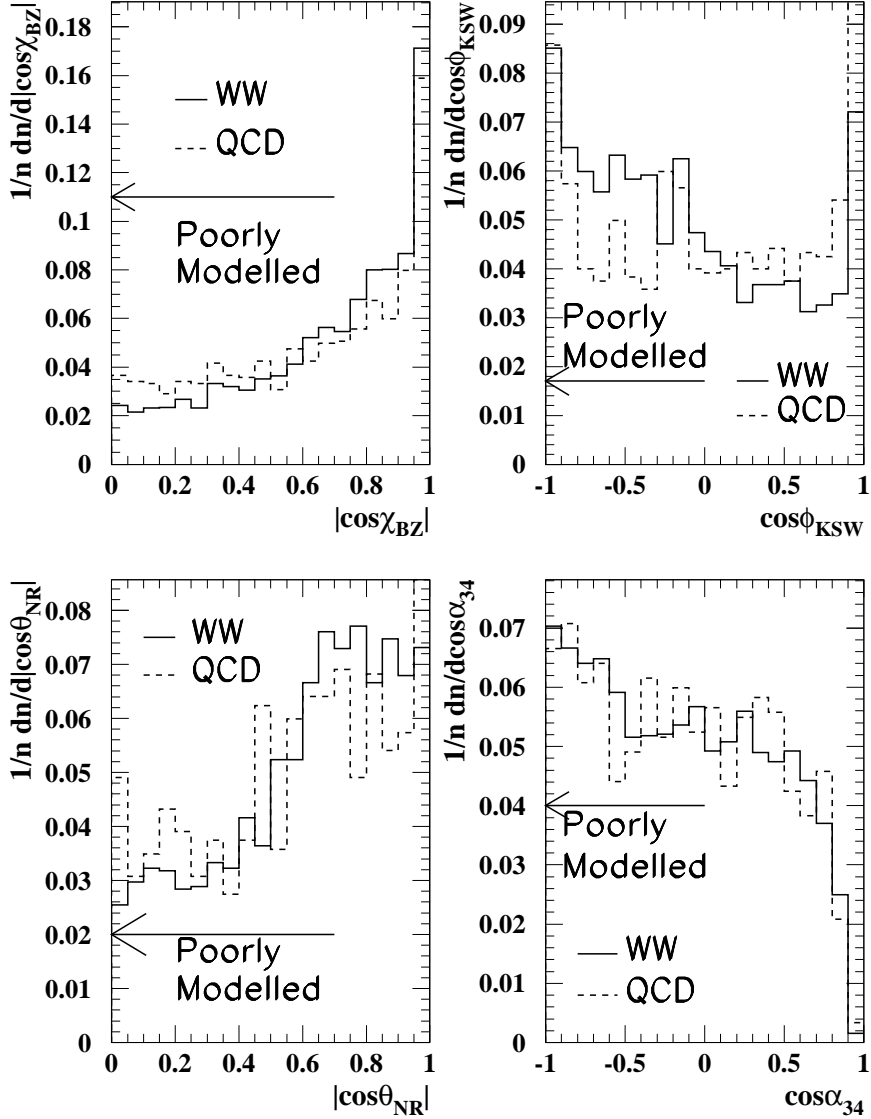}
 \caption{Simulations of  \protect\QCD\ and \protect\WW\ events
at 172~GeV binned according to the four angular variables discussed
in the text}
  \label{fig:4j_rj1}
\end{figure}

Two studies directly addressed this problem at the Workshop. The first 
investigated whether \QCD\ events which pass the \WW\ event selection
do in fact populate
the ranges of the four-jet angular variables where the \MC s are known to have 
problems describing the \LEPONE\ data.
Fig.~\ref{fig:4j_rj1} shows 
simulations of  \QCD\ and \WW\ events
at 172~GeV binned according to the four angular variables listed above.
Two event selections have been used, giving similar results. The first, used
for the workshop itself, was a linear discriminant constructed by \ALEPH\ and
designed for selection of totally hadronic decays of \WW\ pairs at 161 and
172~GeV. Subsequently, this has been replaced by a more generic cuts-based
selection; by energy-scaling the appropriate cuts, this has also been run on
\JETSET\ \PaSh\ Monte Carlo and real data at the Z$^0$ peak, and the discrepancies
observed by \ALEPH\ are seen to persist. The problem regions can be roughly
characterised as follows:
\begin{equation}
\vert \cos\chi_{\mathrm{BZ}} \vert , \vert\cos\theta^*_{\mathrm{NR}}
  \vert  < 0.7,  
\qquad
\vert  \cos\phi_{\mathrm{KSW}}\vert , \vert \cos\alpha_{34} \vert  <  0.5 . 
\end{equation}
The plots shown in Fig.~\ref{fig:4j_rj1} correspond to  generic
cuts: the \QCD\ predictions are  taken from \PYTHIA\ and the \WW\ predictions
are obtained from  \KORALW\ at 172~GeV. 
It is clear that both the selected \WW\ events and the
predicted  \QCD\ background do populate the regions of concern. 
For example,
approximately half the \QCD\ events have
$  \vert \cos\alpha_{34} \vert  <  0.5  $,
a  region where the \PaSh\ \MC s overestimate the \LEPONE\ data  by up
to 15\% \cite{GC}.  

The second study attempted to address the question of the origin 
of the disagreement between the \PaSh\ \MC\ and  matrix element 
predictions for the angular variable distributions. 
 Perturbative \QCD\ predicts very specific angular correlations between
the four final-state partons in $\epem \to \qq\qq, \qqgg$ (see for example
Ref.~\cite{QCDbook}). These correlations are naturally included in a full
matrix element calculation, but are not necessarily included in a \PaSh\ emulation
of the four-jet final state.  

\begin{figure}[tb]
\begin{center}
\mbox{\epsfig{file=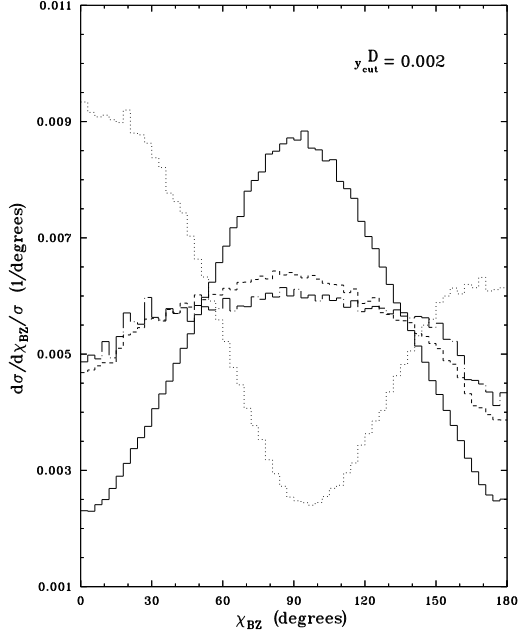,angle=0,height=11cm}}
\caption{Differential distributions in the Bengtsson-Zerwas angle}
\label{fig:4j_sm1}
\end{center}
\end{figure}
\begin{figure}[tb]
\begin{center}
\mbox{\epsfig{file=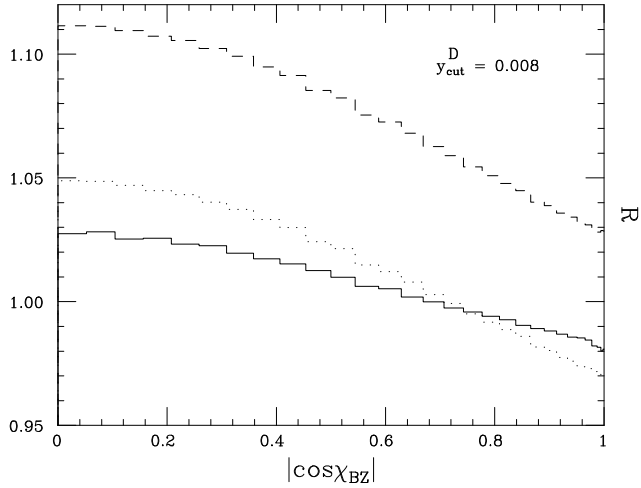,angle=90,height=8cm}}
\caption{The ratio in Eq.~(\protect\ref{ratio})}
\label{fig:4j_sm2}
\end{center}
\end{figure}

To illustrate   these correlations, we show in Fig.~\ref{fig:4j_sm1}
the distributions in $\chi_{\mathrm{BZ}}$ (at \LEPONE, using the Durham
algorithm to define the four-jet sample) calculated for the \QCD\
$\qq\qq$ (solid histogram, exact matrix element) and $\qqgg$ (dotted  
histogram, triple-gluon vertex graphs) processes. 
The former peaks around $90^\circ$,
indicating the preference for the plane of the  two secondary quark jets to be 
orthogonal to the plane of the two  primary  quarks. In contrast, the two 
secondary gluons prefer to be produced in the plane of the primary $\qq$ pair.
Now at \LEPONE\ the $\qqgg$ contribution dominates the total four-jet rate
and so the  $\vert \cos\chi_{\mathrm{BZ}}\vert $ distribution is sharply peaked at 1.
The maximum deviation between the data and the \JETSET\ 
\MC\ occurs for  $\cos \chi_{\mathrm{BZ}} = 0$,
which is precisely the region where the $\qq\qq$ contribution is maximal.
It is possible, therefore, that the \MC\ does not correctly include the 
$\qq\qq$ angular correlations. To study this further, we can construct
a toy matrix element \PaSh-like calculation in which the correlations are 
switched off for the $\qq\qq$ final state, i.e. the secondary $g \to q \bar q$ 
splitting is azimuthally symmetric about the gluon direction. The corresponding
$\chi_{\mathrm{BZ}}$ distribution is shown as the dashed histogram in Fig.~\ref{fig:4j_sm1}.
The distribution is significantly flatter, as expected.  Also shown in the figure
(dot-dashed histogram) is  the prediction of a 
decorrelated version of the $\qqgg$ matrix element (again
only including the triple-gluon-vertex diagrams). 
Since evidently the \MC\ gives
a good description of the \LEPONE\ data for $\vert \cos\chi_{\mathrm{BZ}}\vert \sim 1 $
\cite{GC} we may conclude that the $\qqgg$ correlations are correctly implemented. 
(This is not perhaps surprising, since the `abelianized' $\qqgg$ predictions do
not differ markedly from the \QCD\ predictions \cite{MorTau}.)
Fig.~\ref{fig:4j_sm2} shows the ratio
\begin{equation}\label{ratio}
{\mathrm{R}} = \frac{d\sigma(\qqgg) + d\sigma(\mbox{exact}\; \qq\qq) }
                {d\sigma(\qqgg) + d\sigma(\mbox{decorrelated}\; \qq\qq) }
\end{equation}
as a function of $\vert \cos\chi_{\mathrm{BZ}}\vert $ (solid line). 
The prediction has
the same {\em qualitative} features as the data/\MC\ ratio of the \ALEPH\ analysis \cite{GC}.
This suggests that the lack of correct angular correlations  in the $\qq\qq$
part of the \PaSh\ \MC s  is at least partly responsible for the disagreement
with the \LEPONE\ four-jet data.
However the difference between data and \MC\ seen by \ALEPH\ is quantitatively
larger than the ratio shown in Fig.~\ref{fig:4j_sm2} \cite{GC,ALEPHgluino}.
In fact it appears to be similar in magnitude to what would be obtained by
either switching off the $\qq\qq$ contribution entirely (dashed line
in Fig.~\ref{fig:4j_sm2}) or giving it the same $\chi_{\mathrm{BZ}}$
dependence as the $\qqgg$ contribution (dotted line).
 Further details of this study will be presented
elsewhere \cite{MorSti}.

\newpage

%%%%%%%%%%%%%%%%%%%%%%%%%%%%%%%%%%%%%%%%%%%%%%%%%%%%%%%%%%%%%%%%%%%%%%%%%%
%     text from Pete Renton                    version of 30/7/97
%%%%%%%%%%%%%%%%%%%%%%%%%%%%%%%%%%%%%%%%%%%%%%%%%%%%%%%%%%%%%%%%%%%%%%%%%%

%\setcounter{footnote}{1}
%\section{Experimental issues in the measurement of 
%$M_W$\/\protect\footnotemark[\value{footnote}]}
%\footnotetext[\value{footnote}]{Prepared by ... }
%\label{sub2}
\section{Experimental issues in the measurement of $M_W$}
\label{sub2}

 In this section various aspects of the experimental techniques and problems
in the determination of $\MW$ are discussed. For the 1996 \LEP\ data two
methods were used. These are the measurement of the $\WW$ cross-section
near threshold ($\sqrt{s} = $ 161.3\gev) and the 
direct reconstruction technique for data at $\sqrt{s} = $ 172\gev. 
These two methods are considered in turn. The list of topics addressed
at the workshop is by no means comprehensive.

%\setcounter{footnote}{3}
%\subsection{Direct reconstruction\/\protect\footnotemark[\value{footnote}]}
%\footnotetext[\value{footnote}]{Prepared by P B Renton}

\setcounter{footnote}{3}
\subsection{Threshold method\/\protect\footnotemark[\value{footnote}]}
\footnotetext[\value{footnote}]{Prepared by C J Parkes, 
                                            P B Renton and M F Watson}
\label{subsub12}
The main issue addressed during the workshop was the 
influence of interference between \WW\ production and other four-fermion final
states. This is discussed in sections~\ref{pbr_1}--\ref{pbr_3}.
In section~\ref{pbr_4} the dominant systematic uncertainty resulting from
\QCD\ background, is reviewed.

\subsubsection{Introduction}
\label{pbr_1}
\begin{figure}[hbtp]
 \begin{center}
  \mbox{\epsfig{file=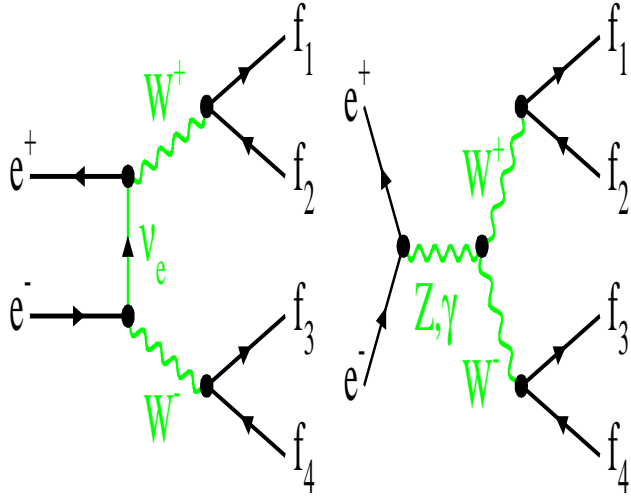,width=.7\textwidth}}
 \end{center}
 \caption[CC03]{Doubly resonant \CCTHREE\ set of diagrams.}
 \label{fig:CC03}
\end{figure}

\begin{figure}[hbtp]
 \begin{center}
 \begin{tabular}{cc}
\mbox{\epsfig{file=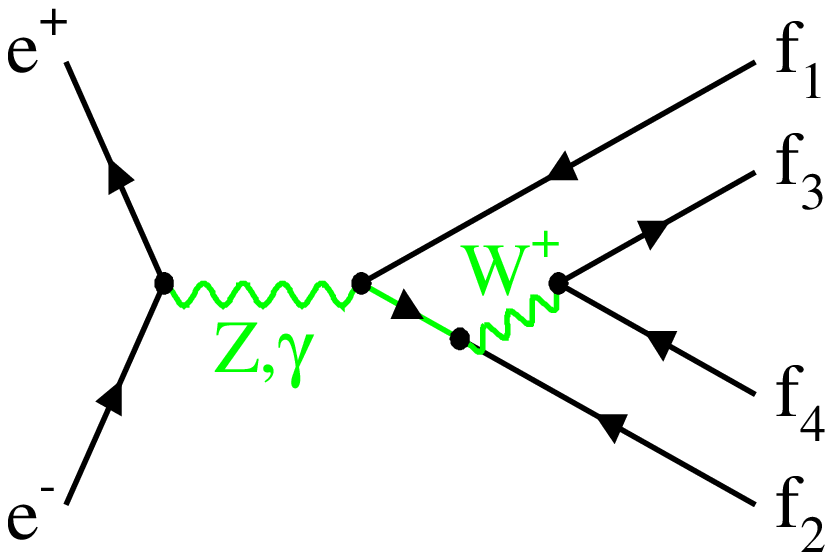,width=.40\textwidth}}
&
  \mbox{\epsfig{file=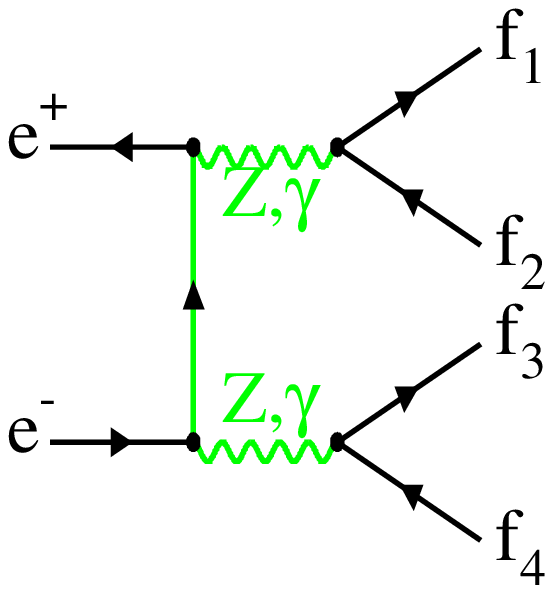,width=.30\textwidth}}
\\
(a) & (b)
\end{tabular}
 \end{center}
\caption{(a) Singly resonant W production,
(b) Additional diagram for
 producing particle anti-particle pairs }
\label{fig:snglzz}
\end{figure}

\begin{figure}[hbtp]
 \begin{center}
  \mbox{\epsfig{file=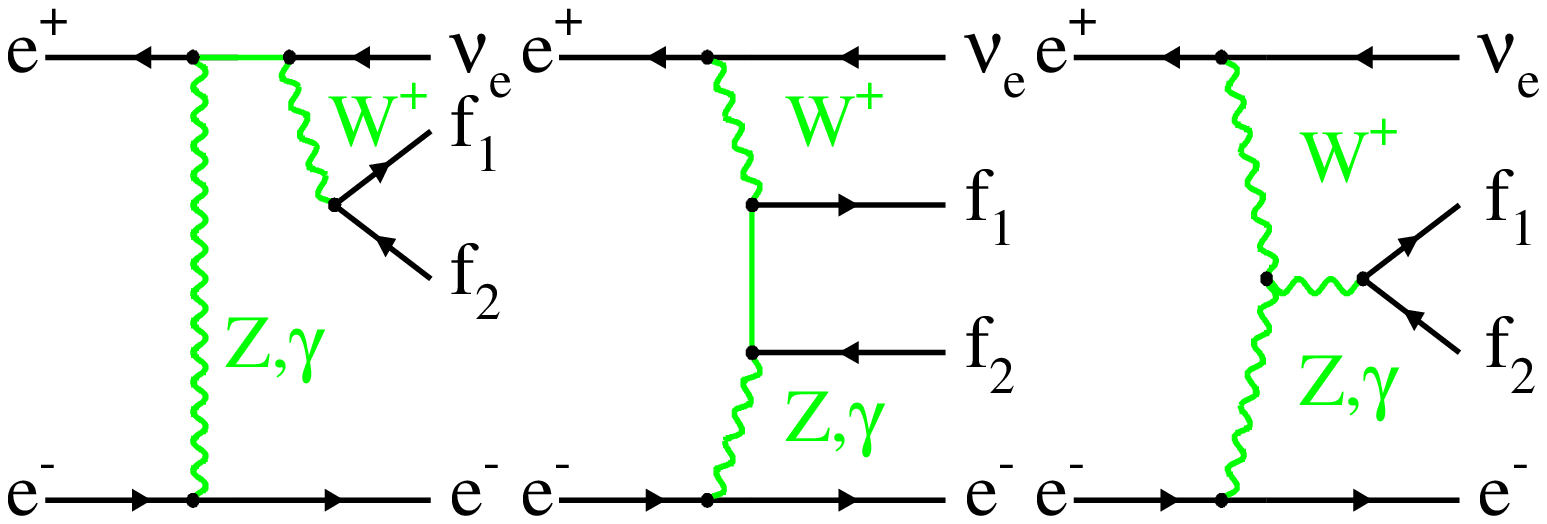,width=.85\textwidth}}
 \end{center}
 \caption[t-chann]
{Example $t$-channel gauge boson graphs for the $\enqq$ channel}
 \label{fig:tchann}
\end{figure}

 The measurement of the $\mathrm{W^+W^-}$ cross-section at \LEP\ is the 
measurement of the
cross-section for the process : 
$\mathrm{e^{+} e^{-} \rightarrow W^{+}W^{-} 
\rightarrow f_{1}f_{2}f_{3}f_{4}}$.
This involves the production of two resonant 
W bosons (via the so-called ``\CCTHREE\ diagrams'' -- see Fig.~\ref{fig:CC03}).
However, identical final states can be produced through different
 intermediate states: 
singly resonant W production (Fig.~\ref{fig:snglzz}(a)), neutral current 
diagrams (Fig.~\ref{fig:snglzz}(b)) and diagrams  containing $t$-channel
 W boson exchange (Fig.~\ref{fig:tchann}), can all contribute (see \cite{YB}). 
Note that the set of diagrams contributing to the
 muon semi-leptonic final state and the tau semi-leptonic channel are 
identical. 
These extra diagrams contribute not only as a background but, as the 
final state is identical to that 
obtained through the \CCTHREE\ diagrams, there is also 
interference {\it between}
these processes: one must sum the matrix element amplitudes, not just consider
the squares of the amplitudes.   

The WW analyses are aimed at selecting events of the type resulting from the
\CCTHREE\ diagrams;
hence the \LEP\ experiments generally choose to interpret their results in terms of 
the  \CCTHREE\ cross-section, rather than the full four-fermion cross-section.
A further significant disadvantage in interpreting the results in terms of
the full four-fermion cross-section is that this quantity is divergent in
several of the Monte-Carlo generators used by the analyses (due to
their massless fermion treatment).
Although the  \CCTHREE\ cross-section is not strictly a gauge invariant quantity
(see \cite{YBTWO}), in practice it is well-defined and 
allows comparison between the measurements made by the experiments.
Clearly, only the doubly resonant \CCTHREE\ subset of diagrams have a 
production cross-section sensitive to the W mass around the threshold energy.
Hence, one does not lose sensitivity in obtaining the W mass from 
the \CCTHREE\ cross-section, rather than from a
four-fermion treatment. 

In this section the methods adopted by the four \LEP\ experiments to 
compensate for the interference effects are considered,
and a quantitative comparison of their results is attempted. 

\subsubsection{Correction method}
\label{pbr_2}
All the experiments used four-fermion generators to produce simulation 
samples for the set of all diagrams and the \CCTHREE\ subset. 
These cross-sections
are defined as $\sigma_{all}$ and $\sigma_{\CCTHREE}$ respectively and
are used to make a correction from the measured (four-fermion)
cross-section to the  \CCTHREE\ cross-section.
 In addition \DELPHI\ have used samples for the non-\CCTHREE\ four-fermion 
diagrams ($\sigma_{4fbck}$).
The selected cross-sections after experimental cuts are denoted 
by $\sigma' = \epsilon \sigma$, where $\epsilon$ is the efficiency.
The following methods were used by the four \LEP\ experiments in their 161\gev\ 
cross-section 
papers~\cite{ALEPH161,DELPHI161,L3161,OPAL161}, where
details of the procedures used can be found.
The aim of all these procedures is to obtain the  \CCTHREE\
cross-section. Here we give the form of the correction term applied by each
experiment: i.e.\ the term which contains the interference effect, as assessed
from Monte-Carlo simulation.

\ALEPH\ and \OPAL\ applied additive corrections to the measured cross-section.
The background from all four-fermion processes is defined to be
  $\sigma'_{all}-\sigma'_{\CCTHREE}$, and has to be subtracted from the
observed cross-section.\footnote{The corrections quoted by \ALEPH\ 
have been divided by the the \CCTHREE\ selection efficiency 
$(\epsilon_{\CCTHREE})^{-1}$
to produce a total, rather than a visible, correction cross-section.} 
The background includes 
a `virtual' contribution due to the interference between
 identical final states; in the $\enqq$\ channel the interference is 
large and negative, so the ultimate sign of the correction could be either 
positive or negative:
in fact in this case the effective background cross-section is negative. 

\DELPHI\ and \LTHREE\ applied multiplicative correction factors.
 For \DELPHI\ this is simply the ratio of the result 
that is obtained when neglecting the interference between
 diagrams to the desired \CCTHREE\ final result, i.e.\
% It is obtained from 
$\sigma'_{\CCTHREE} / \sigma'_{all}-\sigma'_{4fbck}$. 
For \LTHREE\ the correction factor includes the
 total contribution of the four-fermion diagrams and is defined 
as $\sigma_{\CCTHREE}/\sigma'_{all}$. 

\subsubsection{Correction results}
\label{pbr_3}

\begin{figure}[hbtp]
 \begin{center}
  \mbox{\epsfig{file=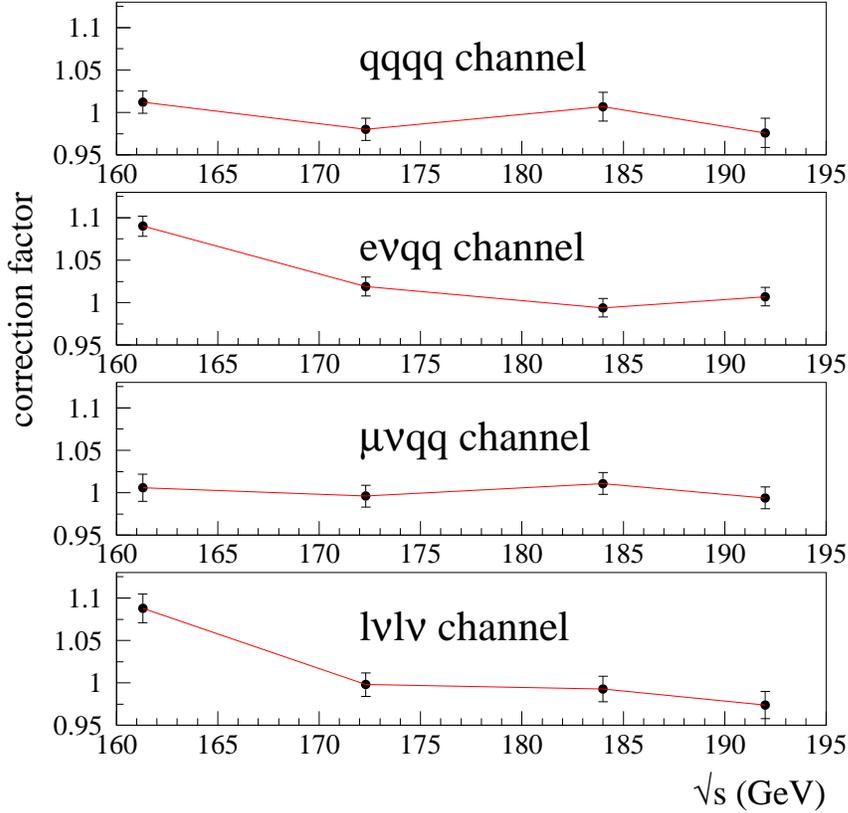,width=.95\textwidth}}
 \end{center}
 \caption[Interference correction factors]{\DELPHI\ interference 
correction factors as a function of the centre of mass energy}
 \label{fig:cor_fac}
\end{figure}

The \DELPHI\ correction factors, as a function of $\sqrt{s}$, are 
given in Fig.~\ref{fig:cor_fac}.
It can be seen that in the $\mnqq$ and $\qqqq$ decay channels no significant
 effects are observed; whereas there are significant effects in 
the $\enqq$\ and $\lnln$\ channels near threshold. The results
were found to remain stable within the statistical 
errors over a wide range of different event selections. 
Thus at 161\gev\ \DELPHI\ finds strong negative interference in 
the $\enqq$\ and $\lnln$\ channels which corresponds to a shift of 
about 30\mev\ in the W mass.

A quantitative comparison of the four-fermion corrections applied by the 
four experiments is complex due to the differences in
procedure. Using information from  
refs~\cite{ALEPH161,DELPHI161,L3161,OPAL161}, 
and other details made available during the workshop, an attempt is made
to put the results of the other experiments in a form similar to the 
published \ALEPH\ numbers.
For the results presented in Table~\ref{tab:res_cor_fac} 
the \OPAL\ and \DELPHI\ values have 
therefore been scaled by the 
inverse of the selection efficiency $(\epsilon_{\CCTHREE})^{-1}$.
There was not sufficient information available to transform 
the \LTHREE\ result to this form.
The \DELPHI\ and \OPAL\ numbers include only the four-fermion 
backgrounds to the 
same final state as is under study, whereas the \ALEPH\ number 
includes in the correction all four-fermion final states.
Whereas the other experiments choose to include the background from 
a Z and radiated gluons in their two fermion production, 
the \ALEPH\ numbers include this in the four-fermion correction: 
this explains the large discrepancy seen in the fully hadronic 
decay channel. Although the \DELPHI\ results show the interference 
effects to be relatively insensitive to the event selection,
when one adds in the full four-fermion backgrounds this will no longer 
be true. Hence the results quoted in Table \ref{tab:res_cor_fac}
will also be sensitive to the analysis cuts.

In general, satisfactory agreement is obtained between the experiments.
However, there could be a possible difference in the results of the two
four-fermion generators used by \OPAL\ .  It is hoped that this study will 
stimulate further co-operation between the experiments on this 
matter. 

\begin{table}[hbt]
\caption[Correction factors  $\sigma'_{all}-\sigma'_{\CCTHREE}$, rescaled 
by the \CCTHREE\ selection efficiency]{Correction
factors  $\sigma'_{all}-\sigma'_{\CCTHREE}$, rescaled by 
the \CCTHREE\ selection efficiency. The full details of the
comparison are given in the text. The $\lnqq$ column is the effective
average of the three leptonic channels.
Note that the result marked $\dagger$ is 
not comparable with the other results in that column
 (as it includes an additional source of background). Numerical results of the
correction factors for \LTHREE\ were not available, but the qualitative
pattern of the corrections is the same as for the other experiments.}
\label{tab:res_cor_fac}
\lineup 
\smallskip
\begin{flushright} 
%\item[]
\footnotesize
\begin{tabular}{llccccc}
\br
Experiment & Generator & \multicolumn{5}{c}{Decay Channel Correction (fb)} \\
           &           & $\qqqq$      & $\mnqq$      & $\enqq$   & $\lnqq$      & $\lnln$ \\ 
\mr
\ALEPH\    & \KORALW\    & +140 $\dagger$ &     &    &$-51\pm$15 & +14  \\
\DELPHI\   & \EXCALIBUR\ & +40$\pm$21  & $-3 \pm$7   & $-34\pm$7  & $-44\pm$15 & +1$\pm$11 \\ 
\OPAL\     & grc4f     & +56$\pm$34 & +6$\pm$17  & $-20\pm$18 & $-26\pm$32   & $-15\pm$18\\
           & \EXCALIBUR\ & +28$\pm$23 & +18$\pm$11 & $-49\pm$12 & $-18\pm$21   & $-21\pm$10\\
\br
\end{tabular}
\normalsize
\end{flushright}
\end {table}  

\subsubsection{Systematics}
\label{pbr_4}

The four experiments agree on the source of the dominant systematic 
error on the WW cross-section: the precision of the
simulation of the $Z(\gamma)$ hadronic decay events. \QCD\ at \LEPTWO\ is 
considered elsewhere in this report 
(\cite{GC} and section~\ref{subsub3}; here the techniques used 
to assess this dominant systematic uncertainty are reviewed.

The background systematics were assessed by the experiments from a comparison 
of events produced by different 
generators/fragmentation packages; for example \OPAL\ compared 
events produced using the \JETSET, \HERWIG\ and \ARIADNE\ models.
 Simulated events were then compared with data, both at 161\gev, and 
because of the limited statistics available, with data
 taken at other energies: for example \DELPHI\ rescaled its selection cuts 
and compared data and simulation at 130-136\gev\
and at 91\gev. The errors ascribed by the experiments 
are given in Table~\ref{tab:sys_QCD}.
The values of the errors assigned are reasonably compatible. The \LTHREE\ value
of this error cannot be compared directly to the others, but it would appear
to be compatible. 
\begin{table}[hbt]
\caption{Systematic error on the threshold cross-section from \QCD\ events}
\begin{indented}
\lineup
\item[]\begin{tabular}{ll}
\br

Experiment & Error \\
\mr
\ALEPH     & 5 $\%$ on background cross-section (data/simulation comparison) \\
           & 6.5 $\%$ on background cross-section (different generators) \\
\DELPHI    & 10 $\%$ on background cross-section \\
\OPAL      & 11.3 $\%$ on background cross-section \\
\LTHREE        & 4 $\%$ on total cross-section \\
\br 
\end{tabular}
\end{indented}
\label{tab:sys_QCD}
\end {table}  

\newpage

\subsection{Direct reconstruction}
\label{subsub11}

A number of topics were discussed at the workshop, and are reviewed below.
In sections~\ref{pbr_5}--\ref{pbr_7} various aspects of kinematic fitting and
mass estimation and their impact in the W mass resolution are discussed.  
In sections~\ref{pbr_8}--\ref{pbr_9} we describe some of the possible biases 
on \MW\ which can occur.   
Then, in sections~\ref{pbr_10}--\ref{pbr_12} the question of four-fermion 
interference effects is discussed.

\setcounter{footnote}{3}
\subsubsection{Kinematic fit methods\/\protect\footnotemark[\value{footnote}]}
\footnotetext[\value{footnote}]{Prepared by P B Renton}
\label{pbr_5}

The resolution on the 
jet-jet mass is too poor to make a precision measurement of $\MW$. 
Fortunately, several methods are available to improve the resolution.  The
methods adopted are either scaling techniques, in which the energy of the
hadronic decay products of the Ws are rescaled to the beam energy, 
or constrained kinematic fitting 
techniques, in which overall energy-momentum conservation is imposed.
These techniques rely essentially on the fact that the jet angles are
measured with good precision, but the jet energies much less so.
Both methods result in a considerable improvement in the jet energy resolution.

 In the absence of additional jets from \QCD\ processes, the 
events WW $\rightarrow \qqqq$ give rise to a four-jet topology. 
The effect of extra
jets is discussed below. If no additional constraints are imposed, 
then the fit for $\Wjjjj$ (where j signifies a jet) 
is a 4-constraint (4-C) fit.
The four fitted jets are combined into two jet-pairs, corresponding to the
two W bosons in the event, and this results in two estimates of $\MW$ per 
event. Three possible choices are possible for this pairing. 
The two measured masses from the 4-C fit
are highly anti-correlated, and can then be combined into a single average
mass using the error matrix from the fit. A second option is to
use a 5-C kinematic fit in which the two fitted masses
for the chosen jet-pairing are required to be equal. This is somewhat 
unphysical, since the masses of the two W's will not, in general, be equal,
but it is found that this procedure gives a good mass resolution.
It is thus of interest to understand which of these choices gives the 
better final precision.

 In the analysis procedures the $\Wjjjj$ event is often first forced into
a 4-jet topology. However, in many events there are additional jets due
to gluon radiation. Treating all events as 4-jet events will result in
some loss of precision on $\MW$. In the preliminary \DELPHI\ analysis events
with 5 jets are also allowed. However, the pairing problem becomes even
more severe, with ten possible combinations.

 The experiments have adopted different methods to tackle the pairing problem
in the fully hadronic channel discussed above. 
\ALEPH\ use a combination of the $\chit$ values from two 
different kinematic fits; a 5-C fit ($\chit_{5C}$) and a fit in which both
masses were compared to a reference mass ($\chit_{ref}$). Using the
combination $\chit_{5C}$ + 2$\chit_{ref}$, Monte Carlo studies show that
the correct solution is chosen in 79\% of the cases. 

 In the \DELPHI\ analysis
all the solutions are retained and used to create an ideogram probability
distribution as a function of $\MW$. Each solution has a Gaussian distribution,
with a width equal to the corresponding 5-C fit error on the mass. The 
relative normalization of the solutions depends on the significance of the
mass difference in the corresponding 4-C fit. Each event is assigned 
a purity $P$
on the basis of the value of the quantity 
$E_{\mathrm{min}}\theta_{\mathrm{min}}$, where
$E_{\mathrm{min}}$ is the minimum jet energy and 
$\theta_{\mathrm{min}}$ the minimum jet
opening angle. This quantity is chosen as it has a rather different shape
for the WW signal events and the Z/$\gamma$ 4-jet background events.
The background in the event is $1-P$, and contains no information on  $\MW$.
The most favoured of the three solutions is normalized to $P$. 
For each solution a likelihood distribution, as a function of $\MW$, is
obtained by convolving a relativistic Breit-Wigner with a Gaussian
distribution for the solution in question, with a width obtained from the
fitted error for that solution. The likelihood distribution for the event
is the sum of these distributions, normalized as described above.
For the events
classified as 5-jets, all 10 combinations are considered in this way.
The probability distributions for each event are then multiplied together
to give an overall likelihood from which the best estimate of $\MW$ and its
error are obtained.

 In the \OPAL\ analysis the 5-C fit with the best $\chit_{5C}$ probability is 
used. This is estimated to contain 68\% of the correct parings. 
In addition, in certain cases, the fit with the 
second best $\chit_{5C}$ is also used.
This is estimated to contain 25\% of the correct parings. The statistical 
problem of retaining more than one solution per event must then of course
be handled.

 The analysis of the semi-leptonic channel is considerably
simpler than that of the fully hadronic channel. The hadronic
decay products are usually forced to form two jets. 
For the $\Wjjln  (\ell = $e,$\mu$) channel there is an undetected neutrino,
so if no additional constraints are imposed then the fit is a 1-C fit.
Imposing equal masses for the leptonic and hadronic decay products
thus gives a 2-C fit.  
For the $\Wjjtn$ there are one or more additional undetected neutrinos, 
so that the $\tau$ lepton energy is poorly known, but the decay 
products can be used as
an estimate of the $\tau$ direction. A rescaling technique can be applied
to the fit results, or directly to the measured jet parameters.
However, the final resolution on $\MW$ is worse in this
channel than for $\ell = $e or $\mu$. 

 The main difference between the experiments in their preliminary analyses of 
the $\Wjjln  (\ell = $e,$\mu$) channel is whether the individual event
errors from the 2-C fit are used. These are used by \DELPHI, but
not so far by \ALEPH\ and \OPAL. A significant difference between the 
$\Wjjln$ and $\Wjjjj$ channels is that the missing neutrino leads to
marked differences in the event topology, and resultant kinematic precision,
depending on its momentum. 

 There is also a variation in the extracted precision coming from the
choice of function used to fit the mass distribution. A study using
\DELPHI\ Monte Carlo data \cite{Julie} has been carried out for 
the $\mnqq$ and $\enqq$ channels and some of the results
are shown in Table \ref{tab:functions}. In each case about 90 samples of Monte
Carlo events, each corresponding to about 10\pb$^{-1}$, were analysed.
From the distribution of the fitted errors the  mean and rms are 
determined. It can be seen that the  use of the event by event errors gives
an improvement of $\simeq$ 15\% in the mean error. The rms spread is
also improved.

\begin{table}
\caption{ Results on the expected error from fits to $\MW$ using various
functional forms and methods. The units are GeV. }
\label{tab:functions} 
\lineup\smallskip
\begin{indented} 
\item[]
\begin{tabular}{lcc} \br
 method   & mean  & rms \\
  \mr
 single Gaussian & 0.598  & 0.197 \\
 single Breit-Wigner & 0.595 & 0.215 \\
 \BW*Gaussian, $\sigma$ free & 0.610 & 0.212 \\
  \BW*Gaussian, event errors & 0.503 & 0.088 \\
\br
\end{tabular}
\end{indented}
\end{table}

\setcounter{footnote}{3}
\subsubsection{\ALEPH\ rescaling technique\/\protect\footnotemark[\value{footnote}]}
\footnotetext[\value{footnote}]{Prepared by J~C~Thompson}
\label{pbr_6}

Preliminary measurements of the W-boson mass from
direct reconstruction have been made for the 1997 Winter Conferences by 
all experiments, and are discussed elsewhere in these proceedings~\cite{JT}. 
All experiments employed a kinematic fit to reconstruct the event invariant 
masses. For the preliminary mass measurements using the 172\gev\ data, 
\DELPHI, \LTHREE\ and \OPAL\ use 5-C kinematic fits to reconstruct the fully 
hadronic final state. 
The \ALEPH\ Collaboration employs the kinematic fit 
in a somewhat different way (referred to here as 4C + Rescaling),
and this is outlined in this section.

 In the \ALEPH\ analysis of hadronic events from the 172\gev\ data, 
events are forced into a 4 jet topology, using the {\scriptsize DURHAM P} 
scheme, 
and a 4-C kinematic fit is applied to 
the energy and direction of each jet constraining the total observed energy 
to be 2$E_{\mathrm{beam}}$ (i.e. ignoring initial state radiation (\ISR)). 
\ALEPH\ Monte Carlo studies have shown 
that, on average, a 4-C fit reproduces the true jet energies more accurately 
than a 5-C fit with equal W masses. Then, to each of the three possible 
combinations of jet-jet pairs, the invariant masses, $m_{12}, m_{34}$, are 
rescaled to obtain a more precise dijet mass. The rescaled mass for jet-jet 
pair (1+2) is defined to be:
\begin{equation}
        m^{R}_{12} = m_{12} \times E_{\mathrm{beam}}/E_{12}
                   = E_{\mathrm{beam}} \times \sqrt{1 - p^{2}_{12}/E^{2}_{12}}
\end{equation}
and similarly for the other pair. The symbols $E$ and $p$ refer to the energy
and momentum of the pair respectively.
Errors coming from the loss of particles, or bad measurement of their 
momenta, are largely cancelled in the ratio $p_{12}/E_{12}$; i.e. the W 
velocities are being used to obtain a reduction of the errors. A simple 
application of 2-body kinematics gives:
\begin{equation}
m^{R}_{12} = m_{12}\left[1 - \frac{m^{2}_{12} 
- m^{2}_{34}}{4E_{\mathrm{beam}}E_{12}}\right]
\end{equation}
demonstrating that the rescaled masses are not improved estimates of 
their respective measured masses, but are now inter-related. It can be 
shown that the correlation is a function of the two true W masses in the 
event, anti-correlated by their measurement errors which are likely to be 
sensitively related to any cuts applied to the two measured di-jet mass 
distributions.

 A sample of sixty events is selected in the data and the best combination
of jet-jet pairs, found as discussed above, is retained. 
The two separate rescaled jet-jet mass distributions are formed 
and each is fitted to a 
simple relativistic Breit-Wigner (\BW), from which $M^{R}_{W1}$,  
$M^{R}_{W2}$ and their respective widths are extracted. $M_W$ is obtained by 
averaging the two extracted masses taking into account the correlation 
expected from the Monte Carlo ($\rho$ = +0.32$\pm$0.10).  

Since the data sample is small, the statistical error is taken from the rms 
spread of the \BW\ fitted masses, $M^{R}_{W1}$,  $M^{R}_{W2}$ obtained 
from 75 Monte Carlo samples each corresponding to 10.6 \pb$^{-1}$, where 
the input mass was 80.5\gev. This gives:
\begin{center}
\begin{tabular}{cccc}
        rms$(M^{R}_{W1})$  &

        rms$(M^{R}_{W2})$  &

        rms$(M^{R}_{ave})$ &

        rms$(M^{5C})$ \\

        463\mev & 483\mev &  420\mev & 506\mev \\
\end{tabular}
\end{center}
where $M^{R}_{ave}$ is the average of  $M^{R}_{W1}$ and $M^{R}_{W2}$, 
taking into account their correlation. Thus the 4C  + 
Rescaling procedure produces a smaller rms by 17\% than the 5C fit.
 The corresponding average fit errors in the 4C + 
Rescaling and 5C analyses from the Monte Carlo samples are 320\mev\ and 
350\mev\ respectively. The final calibration procedure increases these 
rms-based errors slightly: the corrected 
error for the 4C + Rescaling case is 450\mev, which is still smaller than
 the 5C uncorrected fit result.

\setcounter{footnote}{3}
\subsubsection{Further investigation of the rescaling 
technique\/\protect\footnotemark[\value{footnote}]}
\footnotetext[\value{footnote}]{Prepared by M~Thomson}
\label{pbr_7}

The \ALEPH\ rescaling method has been applied to \OPAL\ Monte Carlo data.
Three samples with different input W masses were considered, 78.33\gev,
80.33\gev\ and 82.33\gev. Each was generated at a centre-of-mass energy of
171\gev. Samples with different masses were considered in order to
investigate any possible phase-space effects.

Three different fits to selected fully hadronic final states
were considered, a \mbox{4-C} 
fit where the average mass was used, a 5-C fit, and the
\ALEPH\ rescaling method described above. In each case the jet-pairings which
most closely matched the parton level W pairings were used. No background
was included. One thousand samples of 100 fully simulated events 
were considered.
The rms of the values for $\MW$ obtained from the different methods were taken
as estimates of the error on the fitted mass. The correlation coefficient
between the two masses from the rescaling method was found to 
be $\rho \simeq$ 0.45. Calibration curves
relating the measured to fitted mass were determined (separately for each
fit) from the three different Monte Carlo samples, assuming a linear
relation. The errors on the fitted masses were then scaled appropriately
using the calibration curve to give the error on the measured W mass.
Table \ref{tab:ascale} shows the errors on $\MW$ for the different 
methods and for the different Monte Carlo samples.

\begin{table}
\caption{ Errors obtained using the \OPAL\ analysis procedure for different
kinematical fit procedures, for three values of $\MW$. The errors on $\MW$
are in\mev\, and correspond to a sample of 100 events. }
\label{tab:ascale} 
\lineup\smallskip 
\begin{indented}
\item[]
\begin{tabular}{lccc} \br
type of fit  & $\MW$ = 78.33\gev  & $\MW$ = 80.33\gev & $\MW$ = 82.33\gev
  \\ \mr
 4-C &  369 &  396 & 447            \\
 5-C &  354 &  370 & 339            \\
 rescaled &  362 &  378 & 339            \\
\br
\end{tabular}
\end{indented} 
\end{table}

For the \OPAL\ Monte Carlo samples, the rescaled method performs almost as
well as the 5-C fit. No improvement of $\simeq$ 20\% in the resolution 
is obtained.
It is interesting to note that when phase space effects are less important
(i.e for the $\MW$ =78.33 sample) the performances of three different
kinematic fits are very similar.

In summary, the preliminary \ALEPH\ measurement of $\MW$ from the 172\gev\ data
used 4-C kinematic fitted masses rescaled to the beam energy. 
The \OPAL\ study does not confirm the \ALEPH\ finding that there is an 
improvement of $\simeq$ 20\% in this
rescaling technique compared to the 5-C fit method\footnote{A similar
study by C~Parkes, performed after the workshop using the \DELPHI\ Monte
Carlo data, reached similar conclusions.}.

\setcounter{footnote}{3}
\subsubsection{Systematic shifts on the W mass obtained by direct 
reconstruction\/\protect\footnotemark[\value{footnote}]}
\footnotetext[\value{footnote}]{Prepared by J~J~Ward, A~Moutoussi and R~Edgecock}
\label{pbr_8}

The measurement of the W mass by kinematic reconstruction of the invariant
mass of the W decay products requires several steps. Each of these may be
conceptually well defined, but requires some operational definition,
an estimator, for which there could be various options. The choice is
motivated mainly by achieving high signal efficiency and the best resolution for
the reconstructed mass. However these estimators may be biased, introducing
mass shifts which have to  be evaluated and corrected for using Monte Carlo
events. Such corrections can introduce model dependences and
systematic errors. Various sources of mass shifts are discussed below and the
results are summarised in \Tab{bias}. These studies are carried out using the
\ALEPH\ selection and analysis procedures.

The first step in an  analysis of $\MW$ is the event selection. 
\WW events can be selected using a multivariate analysis (e.g. a
Neural Network) which gives the highest signal to background discrimination. An
upper limit for $\dmw$ of 80\mev\ was calculated, for a given selection 
procedure, as the
difference of the reconstructed mass obtained using the `standard' cut on the
selection variable, from the reconstructed mass obtained applying no such cut.
In addition a `standard' reconstruction analysis
was applied to Monte Carlo events with and without \ISR. The difference in the
results obtained for the two cases was $\approx
350$\mev.

Jet clustering can be performed using different particle
association criteria and combination schemes,
e.g the Jade or the Durham algorithm and  massive
($E$) or massless ($P$) schemes respectively. All types of algorithms
introduce mass shifts. Different association algorithms introduce similar
shifts, but the different combination schemes produce substantially different
results.  The value of $\dmw$ quoted here is the difference between the
results obtained using a $P$ or an $E$ scheme.  $P$ type schemes
associate a larger fraction of particles to the
correct jet, hence also better reproducing the parent-quark direction and
energy. $E$ schemes, which assign masses to the particles, give, on average,
shifts from the parent parton values in energies and angles which
are larger than for the $P$ schemes.
 
The next step of the analysis is forming di-jet pairs. The pairing can be
performed using information related to the di-jet mass itself; such methods
give the highest efficiency for correct pair association but introduce biases
to the measurement. The quoted value of $\dmw$ is the maximum difference
between results obtained using pairing algorithms based on several different
variables, e.g.\ angular separation of the jets, $\chi^2$ of a kinematic fit,
jet charge, etc.

In order to improve the mass resolution a kinematic fit can be applied,
using the constraints of energy and momentum conservation (4-C fit).
In addition, the exact (or Gaussian) equality of the two W
masses of the event can be imposed (5-C fit). The difference in the
reconstructed mass using the two types of fit is given in \Tab{bias}.

Finally, to extract $\MW$ from the reconstructed invariant mass,
the signal is fitted with a function, such as a Breit-Wigner.
However there is no unique form for such a function and, depending on 
the choice, different $\MW$ and errors (from the fit) on $\MW$ can be obtained.
Functional forms used for the fit include convolving 
the Breit-Wigner with a Gaussian
and including a phase space correction factor.
The value $\dmw$ in \Tab{bias} is the maximum difference between results
obtained using different functions and fit ranges; the maximum difference
arises from performing the fit with or without the phase space factor.   
The shifts at generator level from using a phase space factor, or not, 
are about 150 \mev. At the detector level there is additional distortion 
of the line shape from the kinematic fit.

\begin{table}
\caption{{\label{tab:bias}}Breakdown of possible bias components.
  $\dmw$ is the difference between the 
  results for $\MW$ 
  obtained with different analyses 
  (not the difference of the result obtained from the generated $\MW$) 
  A `standard'
  analysis was performed changing every time only the relevant
  component.}
\lineup 
\smallskip
\begin{indented} 
\item[]\begin{tabular}{llc}    \br
Component& Evaluation Method & $\dmw$ \\ \mr
Event selection& Change cut value  &  $\le 80$\mev \\
\ISR         & off$\rightarrow$on              & $\approx 350$\mev \\ 
Jet algorithm & Massive$\rightarrow$Massless scheme & $\approx 350$\mev  \\ 
Jet pairing   & Change algorithm  & $\approx 100$\mev  \\ 
Kinematic fit & 4-C$\rightarrow$5-C  & $\approx 300$\mev   \\ 
\BW\  fit       & Change functional form & $\le 300$\mev \\
\br
\end{tabular}
\end{indented}

\end{table}

As we can see from the \Tab{bias}, the mass shifts and hence
potential biases can be large, but it is the uncertainty
on the necessary corrections that is important for the final measurement.
Hence, at any stage, it is preferable to choose an algorithm that is not only
efficient but also as unbiased as possible, so that any associated errors are
small. Often there are physics grounds on which the choice for the
most unbiased method
can be based. For example, for the jet algorithm one solution is to assign
particles to jets using a P scheme and then recompute the mass of the jet
using an E scheme. Since nearly all the corrections rely on how well
the Monte Carlo events describe the data, it is clearly important that
this agreement is tested as extensively as possible.

\setcounter{footnote}{3}
\subsubsection{Linearity and biases\/\protect\footnotemark[\value{footnote}]}
\footnotetext[\value{footnote}]{Prepared by C~P~Ward}
\label{pbr_9}

 The methods of extracting $\MW$ so far adopted rely heavily on calibration
from Monte Carlo generated samples which are analysed with the same
procedures as the experimental data. In order to ensure that this
calibration does not introduce a bias, Monte Carlo samples for a range
of $\MW$ values are used. The relationship between the fitted 
mass value $\MFIT$ and $\MTRUE$ is determined. This is found to be
essentially linear in the region of interest and can be cast in the form
\begin{equation}
 \MFIT - \rm{80.35} = \aaa ( \MTRUE - \rm{80.35} ) + \bbb \ \ .
\end{equation} 
Ideally one would like the linearity $\aaa =$ 1 and the bias $\bbb =$ 0.
Table \ref{tab:linearity} shows the values of $\aaa$ and $\bbb$ obtained 
for the
preliminary \LEP\ results at the 1997 Winter Conferences.

\begin{table}
\caption{ Available values of the linearity $\aaa$ and bias $\bbb$ for the 
different channels.}
\label{tab:linearity}
\lineup
\smallskip
\begin{indented}
\item[]
\begin{tabular}{lccc} \br
channel & expt. & $\aaa$ & $\bbb$
  \\ \mr
\qqqq & \ALEPH           & 0.938$\pm$0.035 & 0.19$\pm$0.02 \\
     & \DELPHI          & 0.98$\pm$0.07 & $-0.10\pm$0.04 \\
     & \OPAL            & 0.953$\pm$0.012 & 0.095$\pm$0.014 \\
     & \OPAL(no \ISR)    & 0.975$\pm$0.007 & $-0.077\pm$0.008 \\
\mr
$\mnqq$  & \ALEPH       & 0.894$\pm$0.048 & $-0.25\pm$0.03 \\
     & \DELPHI          & 0.80$\pm$0.06 & $-0.25\pm$0.03 \\
     & \OPAL            & 0.89$\pm$0.01 & 0.37$\pm$0.01 \\
\mr
$\enqq$  & \DELPHI      & 0.84$\pm$0.10 & $-0.43\pm$0.05 \\
     & \OPAL            & 0.905$\pm$0.009 & 0.24$\pm$0.01 \\
\mr
$\lnqq$ &  \OPAL        & 0.907$\pm$0.007 & 0.33$\pm$0.01 \\
     & \OPAL(no \ISR)    & 0.940$\pm$0.005 & 0.07$\pm$0.01 \\
\br
\end{tabular}
\end{indented}
\end{table}

 A value of $\aaa$ less than unity results in a corresponding loss of precision
in converting $\MFIT$ to $\MTRUE$. From Table \ref{tab:linearity} it can
be seen that this is most often the case. The \OPAL\ values with and 
without \ISR\ show that
\ISR\ is part of the cause for $\aaa$ being less than unity. Further work by
the experiments is clearly needed to optimize the sensitivity.

\subsubsection{Possible distortion by non-\CCTHREE\ graphs}
\label{pbr_10}

 The presence of  non-\CCTHREE\ graphs will change the shape of the
fermion-fermion invariant mass distribution in the region
around $\MW$ compared to the case of \CCTHREE\ graphs only. Although
this can be taken into account implicitly in the analysis by using
a full four-fermion generator for the ``calibration'' procedure, it
is of interest to understand the size of the effects and the agreement
between different four-fermion generators. Firstly the shifts in a specific
four-fermion generator, \WPHACT, are considered. Then other studies, including more
detailed detector and selection effects, are discussed.

\setcounter{footnote}{4}
\subsubsection{Generator level studies of 
distortion by non-\CCTHREE\ graphs using \WPHACT\/\protect\footnotemark
[\value{footnote}]}
\footnotetext[\value{footnote}]{Prepared by A~Ballestrero}
\label{pbr_11}
 
 An investigation of the shifts between the value of $\MW$ extracted from
a full four-fermion treatment and the sub-set of \CCTHREE\ diagrams has been carried out
at generator level using \WPHACT\cite{WPHACT}. In each case distributions were
generated for  $\MW$ = 80.356\gev\ for the semi-leptonic 
channels $\uden$ and $\udmn$. The W width used 
was $\GW=2.098$\gev. The effects of collinear initial state 
radiation were included, but not final  state radiation. 
Two types of mass distributions
were analysed. First the ``true'' $\qqb$ invariant mass distribution,
using the generated four momenta of these particles, was used. For the second 
distribution the three-momenta of the u,d and $\ell$ were used to compute
the missing momentum {\bf p$_{\mathrm{miss}}$}. This was used as an estimator of
the neutrino three-momentum. The neutrino energy is taken as the
modulus of {\bf p$_{\mathrm{miss}}$}. 
The invariant mass of the $\ell\nbl$ system was
then calculated, and will be referred to as $\MLNREC$. 
%In the absence of \ISR\ this quantity and
%the ``true'' $\ell\nbl$ invariant mass distribution are the same.

\begin{figure}[hbtp]
 \begin{center}
  \mbox{\epsfig{file=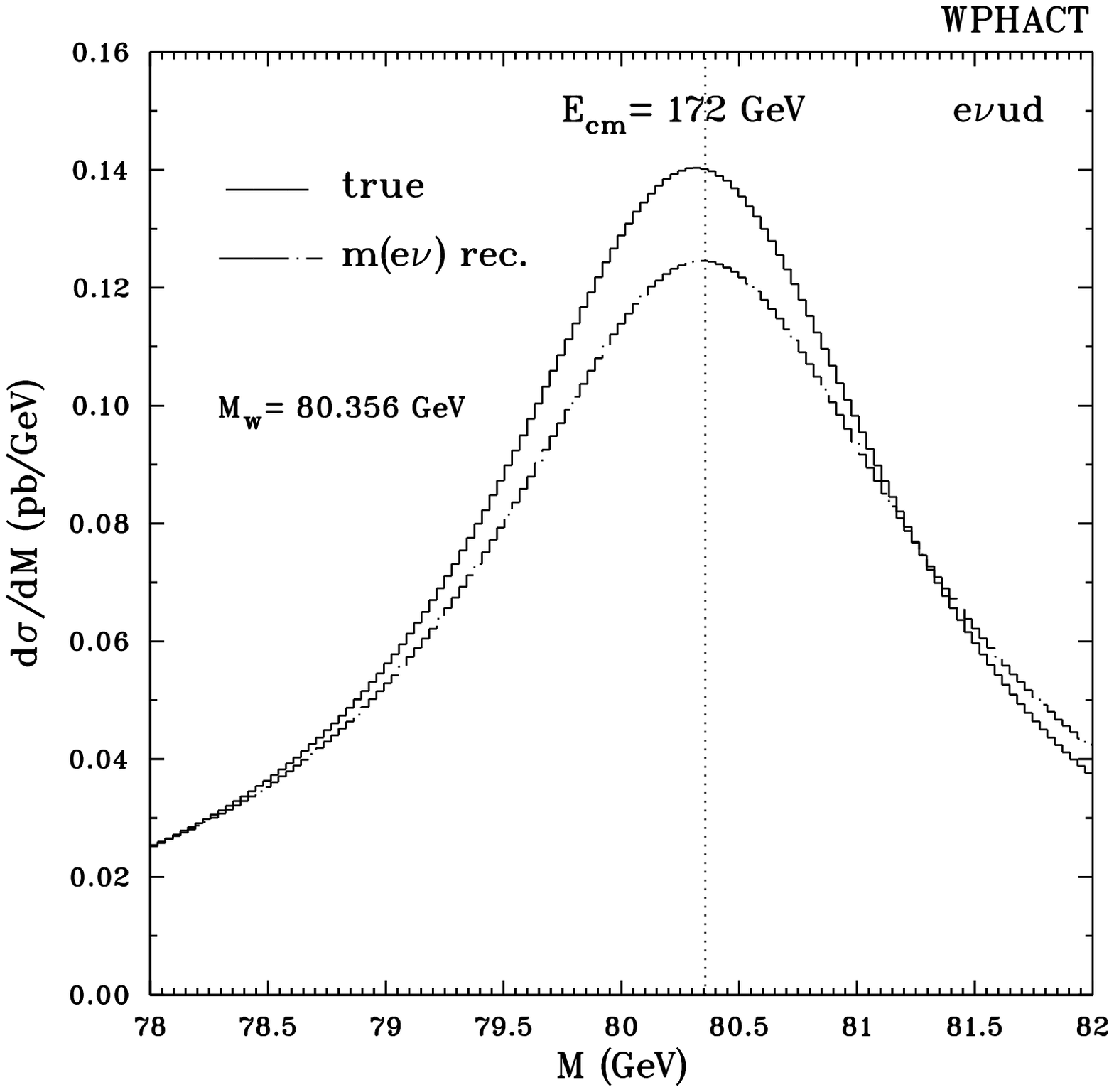,width=.95\textwidth}}
 \end{center}
 \caption[]{Distributions of the ``true'' $\qqb$ 
and ``reconstructed'' $\ell\nbl$ invariant masses from the \WPHACT\ four-fermion
generator for the $\uden$ final state. The definitions of the true $\qqb$
and $\MLNREC$ are given in the text.}
 \label{fig:uden4f}
\end{figure}

 An example of a generated mass distribution for the $\uden$ final state 
is shown 
in Fig.\ref{fig:uden4f} for the full four-fermion case. Some loose event selection 
criteria have been imposed. The minimum of the u, $\overline{\mathrm{d}}$ 
and $\ell$ energies must be 
greater than 10\gev\
and the ud  invariant mass greater than 40\gev. The angle between the lepton
and the beam must be between 10 and 170 degrees, and the $\ell$u 
and $\ell$$\overline{\mathrm{d}}$ angles greater than 5 degrees.

Fits have been made over the mass range shown in the plot to determine the 
W mass and width. A relativistic Breit-Wigner with a running width was used.
The overall $\chit$ for the fits was poor, and the errors  were
multiplied by a factor of two to obtain those quoted below. This gives
an acceptable $\chit$/d.o.f.\ of about unity for the ``true'' distributions,
but is still poor ($\chit$/d.o.f.\ up to 2) for the reconstructed 
distributions. The results are the fits are 
given in Table \ref{tab:WPHACTF}.

\begin{table}
%\caption{ Results of fits to $\MW$ and $\GW$, using the \WPHACT\ generator, 
%to the full four-fermion and \CCTHREE\ diagrams for the ``true'' $\qqb$ and 
%reconstructed $\ell\nbl$ invariant
%mass distributions (see text).
%}
\caption{  Results of fits to $\MW$, using the \WPHACT\ generator,
showing the difference between the full four-fermion and \CCTHREE\ diagrams
 for the ``true'' $\mathrm{\qqb}$ and $\ell\nbl$ masses, and for the
reconstructed $\ell\nbl$ invariant mass distribution (see text). The
units are \mev.}
\label{tab:WPHACTF} 
\lineup 
\smallskip
\begin{indented} 
\item[]
%\begin{tabular}{lcccc} \br
%
% distribution & \multicolumn{2}{c|}{$\uden$} & \multicolumn{2}{c|}{$\udmn$} \\
%        & $\MW$  & $\GW$ &  $\MW$  & $\GW$ \\
%  \hline\hline
%
% 4f $\qqb$ true   & 80.330 $\pm$ 0.001 & 2.095 $\pm$ 0.003 & 80.333 $\pm$ 0.001
% & 2.094 $\pm$ 0.003  \\
%
% 4f $\ell\nbl$ recon.   & 80.388 $\pm$ 0.001 & 2.320 $\pm$ 0.003 & 
% 80.348 $\pm$ 0.001 & 2.337 $\pm$ 0.002  \\
%
%  \hline
% \CCTHREE $\qqb$ true   & 80.333 $\pm$ 0.001 & 2.094 $\pm$ 0.002 & 80.333 $\pm$ 0.001
% & 2.094 $\pm$ 0.002  \\
%
% \CCTHREE $\ell\nbl$ recon.   & 80.348 $\pm$ 0.001 & 2.337 $\pm$ 0.002 &
% 80.348 $\pm$ 0.001 & 2.337 $\pm$ 0.002  \\
%\hline\hline
%\end{tabular}
%\end{indented}

\begin{tabular}{lcc} \br

 distribution       & $\uden$     & $\udmn$   \\ \hline\hline
 $\qqb$ true        &  $-3\pm$1  & 0$\pm$1 \\ \hline
 $\ell\nbl$ true     & +41$\pm$1 & 0$\pm$1 \\ \hline
 $\ell\nbl$ recon.   & +40$\pm$1 & 0$\pm$1 \\
  \br
\end{tabular}

\end{indented} 
\end{table}

 It can be seen from Table \ref{tab:WPHACTF} that the differences in the
fitted $\MW$ values between the full four-fermion and \CCTHREE\ cases for the 
``true'' $\mathrm{\qqb}$ distributions are small. For
the channel $\uden$ the shift is about 3\mev, whereas for $\udmn$ it
is even smaller. The shift with respect to the generated mass is 
about 25\mev, and on the width the change is 4\mev\ or less. 
The ``true'' $\ell\nbl$ fitted masses show no difference in the $\udmn$ 
channel, but a 40\mev\ shift is seen in the $\uden$ channel. 
When the fits are performed on the reconstructed $\ell\nbl$ distributions
a similar pattern is observed: a shift is still found in the $\uden$
channel, and no mass difference occurs in the $\udmn$ channel. 
 The fitted widths also change significantly (increasing by 0.2\gev),
indicating that the shape of the distribution is distorted by the reconstruction
procedure. This is also apparent from the poorer $\chit$ of these fits.

%The fits to the reconstructed $\ell\nu$ distributions give a somewhat
%larger shift, of about 40\mev, for the $\uden$ channel. However,
%the corresponding shift in the $\udmn$ channel is small.
% The fitted widths also change significantly, indicating
%that the shape of the distribution is distorted by the reconstruction
%procedure. This is also apparent from the poorer $\chit$ of these fits.

\setcounter{footnote}{3}
\subsubsection{Other studies of distortion 
by non-\CCTHREE\ graphs\/\protect\footnotemark
[\value{footnote}]}
\footnotetext[\value{footnote}]{Prepared by C Parkes, P B Renton and D Ward}
\label{pbr_12}

This section reports on studies performed by the \OPAL\ and \DELPHI\
Collaborations comparing \CCTHREE\ and full four-fermion simulated events.

In the \OPAL\ study simulation events corresponding to an 
integrated luminosity of 
around  5000 \pb$^{-1}$ were generated. The full \OPAL\ detector simulation and 
reconstruction methods were then applied. Events which pass the 
usual selection criteria
are then analysed in two complementary ways.

The first method applied by \OPAL\ uses the true generated W mass $\MTRUE$
for each selected event. $\MTRUE$ is calculated by appropriately pairing up the
generated four final state fermions as if they came from $\WW$.
%This procedure is, of course, not exact for the fully hadronic
%channel. 
Two invariant  masses are then formed, and the average is
found. A fit is made to the distribution of $\MTRUE$  to a 
function BW($M)\times p(M)$,  where BW($M$) 
is a relativistic Breit-Wigner, and $p(M)$ 
is a phase-space factor equal to the c.m. momentum of a pair of W's of 
average mass $M$.
Then a comparison between the fitted mass parameter in the \BW\ for \CCTHREE\ 
and four-fermion cases is made, and the results for (four-fermion $-$ \CCTHREE) 
(in\mev) are shown 
in Table \ref{tab:cc034f}
for the different channels and for two four-fermion generators; grc4f and 
\EXCALIBUR.

%     A comparison has been made between CC03 and full four-fermion simulations
%to study such effects. This has been done using two approaches.

%In the first analysis simulations made at 171\gev\ are used,
%with samples corresponding to a luminosity of around 5000 \pb$^{-1}$. 
%The full \OPAL\ detector simulation and reconstruction methods are used in 
%these studies.
  
% The procedure is as follows. Events which pass the usual selection
%criteria are retained for further study. These are then analysed in two
%complementary ways.

% In the first analysis the {\it true W mass, $\MTRUE$} is calculated for  each 
%selected event. This is done by finding the four final state fermions 
%as generated, and then pairing them up appropriately  as if they came 
%from $\WW$. This procedure is, of course, not exact for the fully hadronic
%channel.Two invariant  masses are then formed, and the average is
%found. A fit is made to the distribution of $\MTRUE$  to a 
%function BW(M)*P(M),  where BW(M) is a relativistic Breit-Wigner, and P(M) 
%is a phase-space factor equal to the c.m. momentum of a pair of W's of 
%average mass M.
%Then a comparison between the fitted mass parameter in the BW for CC03 
%and four-fermion is made, and the results for (4f-CC03) (in\mev) are shown 
%in Table \ref{tab:cc034f}
%for the different channels and for two four-fermion generators; grc4f and 
%\EXCALIBUR.

\begin{table}
\caption{ Differences between the values of $\MTRUE$ between four-fermion 
and \CCTHREE,
for different channels and two different four-fermion generators. 
The units are\mev. }
\label{tab:cc034f}
\lineup
\smallskip
\begin{indented}
\item[]
\begin{tabular}{lccc} \br
generator  & $\qqqqa$  & $\qqen$ & $\qqmn$($\qqtn$)
  \\ \mr
 grc4f  &  $-28\pm$15 & +26$\pm$23 & $-23\pm$18 \\
 \EXCALIBUR\  &  $-4\pm$13 & +58$\pm$19 & +3$\pm$15 \\
\br
\end{tabular}
\end{indented}
\end{table}

Despite the large statistics in these samples, the results are still somewhat
inconclusive.  Both generators seem to indicate
a positive mass shift for $\qqen$ compared to the other channels, at the 
level of about 50\mev, albeit with large errors. These results confirm the naive
 expectation that the shift is larger in the $\qqen$ channel than the 
$\qqmn$ channel, as additional diagrams contribute in the electron
 channel (see section 3.1.1)
However, there also seems to be an overall shift of the grc4f 
numbers with respect to the \EXCALIBUR\ numbers, by about 25\mev. This,
of course, is not expected.

 In the second \OPAL\ analysis the difference $\MREC$ - $\MTRUE$, where $\MREC$  
is the reconstructed mass, is formed. From a fit to the peak region of this
difference the mean is found. This is then a measure of the 
bias introduced by the reconstruction.  Next the difference in this bias
between four-fermion and \CCTHREE\ is found. This study is done for events with only 
a small amount of \ISR, because \ISR\ distorts the resolution peak.
The results are given in Table \ref{tab:cc034fa}. 

\begin{table}
\caption{ Mean of the differences between the values 
of  $\MREC$ - $\MTRUE$ between four-fermion and \CCTHREE,
for different channels and two different four-fermion generators. 
The units are\mev. }
\label{tab:cc034fa}
\lineup
\smallskip
\begin{indented}
\item[]
\begin{tabular}{lccc} \br
generator  & $\qqqqa$  & $\qqen$ & $\qqmn$($\qqtn$)
  \\ \mr
 grc4f      &  0$\pm$12    & $-26\pm$20 & $-30\pm$20 \\
 \EXCALIBUR\  &  +23$\pm$10 & $-5\pm$16  & +28$\pm$16 \\
\br
\end{tabular}
\end{indented}
\end{table}

 It is again difficult to draw very clear conclusions from this set 
of numbers. Naively one would have expected these all to be close to zero;
and indeed they may, but only two lie within 1 s.d.\ of zero.
Again, the grc4f numbers are systematically shifted compared to \EXCALIBUR. 
Whether these differences stem from the original generators or from the
modifications needed to implement them in the experimental
environment (e.g.\ Coulomb corrections, \QCD\ corrections to $\GW$, an angle cut 
on the electron, treatment of masses etc.) is still an open question.

 The \DELPHI\ study was made using the \EXCALIBUR\ four-fermion 
generator to produce \CCTHREE\
and four-fermion event samples. The event selection was made at 
the generator level, where
cuts were applied that mimic those of the collaboration's full analysis. 
The samples
were then analysed in a similar manner to the first \OPAL\ method. 
The results of
this study, and a further similar study in 
the \DELPHI\ Collaboration \cite{KMONIG},
are compatible with the \OPAL\ \EXCALIBUR\ results given 
in table \ref{tab:cc034fa}.
The shifts in the $\qqqq$ and $\mnqq$ channels are found to be small 
( $\lapproxeq$ 10\mev). For the $\enqq$ channel the shifts are small
( $\lapproxeq$ 20\mev) when fitting the $\mathrm{\qqb}$ invariant mass, but about 75
\mev\ when fitting the $\ell\nbl$ invariant mass.
Thus it would appear that it is mainly in the $\ell\nbl$ invariant mass
distribution that a distortion of the Breit-Wigner shape by non-\CCTHREE\
graphs occurs.
% \DELPHI\ produced results using a fast detector sim.

% A study has also been made using the \EXCALIBUR\ four-fermion generator and using
%cuts at the generator level which mimic those used by \DELPHI, for both
%sample of all diagrams and the CC03 subset.
%For the $\mnqq$ channel the four-fermion to CC03 shifts are compatible with zero.
%For the $\enqq$ channel there is a shift of 70 $\pm$ 20 (???)\mev\ when
%fitting the ``true'' $\ell\nu$ invariant mass distribution over the
%range x to y\mev. The corresponding shift for a fit to the $\qqb$
%distribution is 30 $\pm$ 20\mev. A further similar study in the
%\DELPHI\ Collaboration \cite{KMONIG} gives shifts in the $\mnqq$ and
%$\qqqq$ channels which are small ( $\lapproxeq$ 10\mev). For the $\enqq$
%channel the shifts are small (10\mev\ or less) when fitting the $\qqb$
%invariant masses, but about 50\mev\ when fitting the $\ell\nu$ invariant mass.
%Thus is would appear that it is mainly in the $\ell\nu$ invariant mass
%distribution that a distortion of the Breit-Wigner shape by non-CC03
%graphs occurs.

\newpage 

\setcounter{footnote}{3}

\subsection{Error from the determination of the \LEP\ 
energy\/\protect\footnotemark[\value{footnote}]}
\footnotetext[\value{footnote}]{Prepared by P~B~Renton}
 At \LEPONE\ the absolute energy scale is determined by the
very precise method of {\it resonant depolarisation}, which gives
an accuracy of better than 1\mev\ in the average circulating beam energy.
This technique becomes increasingly difficult at higher beam energies,
so that the accuracy at high energy is largely determined by the highest
value of the energy at which resonant depolarisation can be made.

 Several attempts were made in 1996 to  achieve transverse polarisation
at as high an energy as possible. A measurable level of polarisation was
obtained at 50\gev\ on several occasions, but so far attempts at higher
beam energies have not succeeded.

 The 1996 beam energy was obtained with a precision of 
about 30\mev~\cite{LEPEN}.
By far the largest component of the error came from extrapolating the
polarisation measurements at lower energies (45 and 50\gev) to the operating
beam energies of 80.5 and 86\gev. The extrapolation errors were 24 and 29
\mev\ respectively. The other significant components in the error came from
the relative fill to fill normalization (10 and 5\mev\ respectively) and
the uncertainties in the modelling of the RF corrections (5 and 5\mev\
respectively). The \LEP\ beam energy error is a common systematic
error when combining results from the four \LEP\ experiments. However,
because of the limited luminosity for the 1996 data, the \LEP\ energy
error remained a small component of the overall $\MW$ error.

 In order that the error from the \LEP\ beam energy remains a small component
in future high luminosity running it is desirable that it can be reduced 
to 15\mev\ or better\footnote{In
\cite{YB} a target error of 12\mev\ was used}. In order to achieve this the
extrapolation error must be reduced. Two approaches are being pursued by the
\LEP\ Energy Working Group. The first is to understand better the relative
energy scales at 45 and 50\gev, where measurements have been made. The relative
error ($\sigma_{\rm{E}}$) for these two energies in 1996 was about 4\mev; 
and this leads to the extrapolation errors quoted above. A significant
part of this error comes from the rather limited number of such polarisation
measurements. In only one fill was the energy measured at both these energies,
and in addition there were two fills where measurements at 50\gev\ were made.
More depolarisation measurements at these
energies should, provided there are no surprises, allow a significant reduction
in $\sigma_{\rm{E}}$. 

 The extrapolation with energy in 1996 was taken to be linear. 
If $\sigma_{\rm{E}}$ can be substantially reduced, then possible 
non-linear terms in the extrapolation could then constitute a 
sizeable part of the extrapolation error.
Thus obtaining resonant depolarisation at energies higher than 50\gev\ is
a high priority. During the 1996/97 shut-down improvements were made to \LEP\
which are designed to produce flatter orbits, and thus improve the polarisation
build up at higher energies. Polarisation measurements at 55\gev, or possibly
60\gev, would help considerably in understanding the extrapolation.
If polarisation at higher energies cannot be achieved then there is a 
possibility to get a third point somewhat below 45\gev. However this 
possibility is limited by the range of the \NMR\ devices which measure the 
dipole field and are needed in the extrapolation and modelling procedure.

 In summary, provided that there are no hidden surprises, and that sufficient
time is devoted to making resonant depolarisation measurements, then a \LEP\
beam energy error of 15\mev\ should be achievable.

\newpage

\setcounter{footnote}{3}
\subsection{Error extrapolation\/\protect\footnotemark[\value{footnote}]}
\footnotetext[\value{footnote}]{Prepared by D~G~Charlton,  P~Dornan,
 R~Edgecock,  P~B~Renton,  W~J~Stirling, M~F~Watson}
\label{subsub13}

% signif alterations follow labels % CJP

The report of the \MW\ Working Group in the \LEPTWO\ Workshop 
Yellow Book (YB) \cite{YB}  
contains estimates for the likely attainable precision on \MW\,
from both the threshold and direct reconstruction method.
% CJP
These estimates contain the most accurate information available 
in late 1995; they were based on Monte Carlo simulation work and
theoretical studies.
However, now that a significant amount of \LEPTWO\ data has been 
collected and 
\MW\ measurements have been made using both methods 
(for a summary see \cite{JT}),
it is worth reviewing these  estimates.
In the following sections we compare the precision currently achieved
by the LEP collaborations with that predicted in the Yellow Book study, and use this
 comparison to refine the predictions for future attainable precision
on \MW\ as a function of luminosity.

\subsubsection{Threshold method}
% CJP
The method of extracting \MW\ from the \WW\ cross-section at threshold
 is discussed in \cite{JT}. In 1996 the four LEP experiments combined 
obtained a total luminosity of approximately $40\pb^{-1}$ at a centre of
mass energy of $161\gev$. In Tables \ref{tab:lnln}-\ref{tab:qqqq} we consider
the three decay channels ($\lnln$, $\qqln$, $\qqqq$) and compare the Yellow 
Book predictions (labelled YB) with the published experimental results 
of the four \LEP\ collaborations (labelled achieved); in the light of experience
we then provide new estimates that would be obtainable for future running.  
It is important to note that great effort has not yet been made in reducing
the systematic errors, since for the current integrated luminosity the
statistical errors are completely dominant: indeed currently the statistical errors
from the number of simulation events produced can be a significant component
 of the systematic error.

Table \ref{tab:lnln} shows the signal efficiency, background cross section, 
and systematic error
for the $\lnln$ decay channel. 
The systematic errors obtained by the four experiments already largely
 match the YB estimates. In addition, \ALEPH\ and \OPAL\ report 
 combined efficiencies for this channel
 in excess of the $60$\% YB prediction, however the backgrounds
are somewhat larger than anticipated. To extrapolate to 
 higher luminosities, we assume an efficiency
 of $\epsilon = 75\%$  and a corresponding background estimate
 of $0.02\pb$. We assume a (common) systematic error of $4\%$,
 a value  already achieved
 by one experiment. Using these values we may estimate the 
 fractional errors on $\sigma_{\rm WW}$ for different luminosities.
 Table~\ref{tab:overall} compares the estimated (`Oxford') and YB errors
 for ${\cal L} = 10, 50, 100\pb^{-1}$ per experiment. Note that
 for this channel the Oxford and YB estimates are almost identical, the 
 increase in background being compensated for by a slightly
 higher  efficiency. 
\begin{table}[htb]
\caption{Signal efficiency, background cross section, and systematic error
for the $\lnln$ channel.} 
\label{tab:lnln} 
\lineup 
\smallskip
\begin{indented} 
\item[]\begin{tabular}{|l|c|c|c|c|} \hline
 & $\sigma_{\rm sig}$  & $\sigma_{\rm bkg}$ & $\epsilon$  &   
 $(\Delta\sigma/\sigma)_{\rm sys}$  \\ \hline
 YB &  $0.23\pb$ & $0.01\pb$ & $60$\% & $2.7$\%           \\
 Achieved &   & $0.02 - 0.06\pb$ & $40 - 75$\% & $4-30$\%           \\
 Projection &   & $\sim 0.02\pb$ & $\sim 75$\% & $\lapproxeq 4$\%           \\
  &   &  &  & (mostly common)          \\
\hline
\end{tabular}
\end{indented} 
\end{table}
 
The results of the $\lnqq$ decay channel study are given in
 Table \ref{tab:qqln}.
The most significant difference between the results achieved 
and the YB predictions is in the substantial
improvement in efficiency in the  $\qqtn$ channel.
The YB study made no attempt to optimize this channel, and as a result
the efficiency was only $5\%$. In contrast, the \LEP\ experiments
have already achieved $40 - 50\%$ efficiency, albeit with 
a significantly larger background. For the projection
to higher luminosities we assume an overall efficiency
 of $\epsilon = 75\%$  and a corresponding background 
 of $0.08\pb$. We assume an overall  systematic error of $3\%$
 for this channel, of which $2\%$ is taken to be common.
Comparing the fractional errors on the $\qqln$ cross section listed in
 Table~\ref{tab:overall}, we see a significant improvement
 in the predicted overall precision with respect to the YB.
\begin{table}[htb]
\caption{Signal efficiency, background cross section, and systematic error
for the $\qqln$ channel.} 
\label{tab:qqln} 
\lineup 
\smallskip
\begin{indented} 
\item[]\begin{tabular}{|l|c|c|c|c|} \hline
 & $\sigma_{\rm sig}$  & $\sigma_{\rm bkg}$ & $\epsilon$  &   
 $(\Delta\sigma/\sigma)_{\rm sys}$  \\ \hline
 YB &  $0.76\pb$ & $0.03\pb$ & $47$\% & $2.7$\%           \\
 Achieved &   & $0.08 - 0.19\pb$ & $60 - 76$\% & $3-10$\%           \\
 Projection &   & $\sim 0.08\pb$ & $\sim 75$\% & $\sim 3$\%           \\
  &   &  &  & ( $\sim 2$\% common)          \\
\hline
\end{tabular}
\end{indented} 
\end{table}
 
It is more difficult to make a similar assessment  for the
expected precision for the $\qqqq$ channel. After a loose
preselection to eliminate the major component of the 
$\qqb(\gamma)$ background,
the four experiments use a combination of multivariate and sophisticated 
cut based
analyses to separate the $\WW$ signal from the QCD background.
In contrast, the YB predictions were based on a simple cuts selection \cite{YB}, 
yielding a signal efficiency of approximately $55\%$ and a purity
of approximately $70\%$.
% CJP
Given the nature of the analyses performed, instead of quoting backgrounds
 and efficiencies, we provide a statistical sensitivity both for the
present results and for our predictions. 
   In order to extrapolate to higher luminosities
we take the current \ALEPH\ statistical sensitivity of $\pm 28\%$ 
for $10\pb^{-1}$ as typical, and assume a common systematic error
of $5\%$, slightly larger than the YB estimate, see 
Table \ref{tab:qqqq}.  
\begin{table}[htb]
\caption{Signal efficiency, background cross section, and 
statistical/systematic error
for the $\qqqq$ channel.} 
\label{tab:qqqq} 
\lineup 
\smallskip
\begin{indented} 
\item[]\begin{tabular}{|l|c|c|c|c|c|} \hline
 & $\sigma_{\rm sig}$  & $\sigma_{\rm bkg}$ & $\epsilon$  
& $(\Delta\sigma/\sigma)_{\rm stat}$   & 
  $(\Delta\sigma/\sigma)_{\rm sys}$  \\ \hline
 YB &  $0.94\pb$ & $0.39\pb$ & $55$\% & & $4$\%           \\
 Achieved &   &  &  & 28\%  (\ALEPH )  &         \\
 Projection &   & & & $\sim 28$\% & $\sim 5$\%           \\
  &   &  & & & (common)          \\
\hline
\end{tabular}
\end{indented} 
\end{table}
The corresponding fractional errors on $\sigma_{\rm WW}$
are listed in  Table~\ref{tab:overall}. The improved
signal selection efficiency leads to an overall slight reduction
in the cross section error compared to the YB estimates.
As anticipated in \cite{YB}, similar errors on the cross-section 
are obtained in the $\qqln$ and $\qqqq$ channels.

In order to translate a cross section error into an error
on $\MW$ several additional (common) systematic uncertainties need to
be taken into account.  These are summarised in Table~\ref{tab:common}.
As discussed in detail in Section~3.3, it is estimated that
the \LEP\ beam energy error (currently $\pm 30\mev$) can ultimately
be reduced to $\pm 15\mev$. A value of $\pm 12\mev$ was assumed in the
YB study \cite{YB}.  
The theoretical error on the $\WW$ cross section also contributes
to the error on $\MW$: $\Delta \MW \approx 17\mev\times
(\Delta\sigma/\sigma)_{\rm thy}$. In the YB study a $\pm 2\%$ error
was assumed, arising in part from (unknown)  higher-order loop diagrams involving
Higgs boson exchange. However calculations performed since the 
$\LEPTWO$  workshop \cite{BVO}
 have shown that the corrections are 
intrinsically small and decrease rapidly with increasing \mbox{$M_{\mathrm{H}}$}.
 Using
the range of values for  obtained \mbox{$M_{\mathrm{H}}$} from direct searches and
 global electroweak fits essentially eliminates the error from this source.
The remaining theoretical error on $\sigma_{\rm WW}$ is therefore expected
to come from higher-order finite QED corrections, and so it is reasonable
 to assume an error of $\pm 1\%$ for future projections.

\begin{table}[htb]
\caption{Additional common systematic errors for the 
threshold mass measurement. Note that $\Delta \MW \approx 
\Delta E_{\rm beam}$ and $\Delta \MW \approx 17\mev\times
(\Delta\sigma/\sigma)_{\rm thy}$.} 
\label{tab:common} 
\lineup 
\smallskip
\begin{indented} 
\item[]\begin{tabular}{|l|c|c|c|c|} \hline
 & $\Delta E_{\rm beam}$  &    $(\Delta\sigma/\sigma)_{\rm thy}$  \\ \hline
 YB & $12\mev$  & $2\%\ (\to \ 34\mev)$         \\
 Achieved &   $30\mev$  & $2\%\ (\to \ 34\mev)$     \\
 Projection &   $15\mev$  & $1\%\ (\to \ 17\mev)$    \\
\hline
\end{tabular}
\end{indented} 
\end{table}

Combining these two systematic errors with the cross section errors
discussed above gives the estimated overall errors on $\MW$
at ${\cal L} = 10, 50, 100\pb^{-1}$ (per experiment) listed in Table~\ref{tab:overall}.
Also shown are the YB estimates. (It is interesting to note that the YB 
estimate of $\pm 220\mev$ for $10\pb^{-1}$ is exactly the current error
from the combined \LEP\ experiments.) The key point to note is  that 
a significant improvement on the projected YB errors can be expected. 
In particular, a relatively modest ${\cal L} = 50\pb^{-1}$ total
luminosity per experiment could now  be expected to yield an error of
$\Delta \MW = 86\mev$ from the threshold cross section measurement.

 \begin{table}[htb]
\caption{Anticipated and new projected fractional errors on
$\sigma_{\rm WW}$ and corresponding overall combined error on $\MW$
for three different  total luminosities.}
\label{tab:overall} 
\lineup 
\smallskip
\begin{indented} 
\item[]\begin{tabular}{|l|l|c|c|} \hline
luminosity/expt. & channel      &   YB  & Oxford  \\ \hline
 $10\pb^{-1}$    & $\lnln $  &   34\%  &   32\%     \\
                 & $ \qqln $  &   19\%  &   15\%     \\
                 & $ \qqqq$  &   20\%  &   15\%     \\
                 & combined     &   13\%  &   10\%     \\
     & $\Delta\MW$ (total)     &  {\bf 220 MeV}  &   {\bf 170 MeV}   \\ \hline
 $50\pb^{-1}$    & $\lnln$  &   15\%  &   15\%     \\
                 & $\qqln$  &    9\%  &    7\%     \\
                 & $ \qqqq$  &   10\%  &    8\%     \\
                 & combined     &  5.9\%  &  4.9\%     \\
     & $\Delta\MW$ (total)     &   {\bf 108 MeV} &  {\bf 86 MeV}  \\ \hline
 $100\pb^{-1}$   & $\lnln$  &   11\%  &   11\%     \\
                 & $ \qqln$  &  6.5\%  &  5.1\%     \\
                 & $ \qqqq $  &  7.3\%  &  6.7\%     \\
                 & combined     &  4.4\%  &  3.8\%     \\
     & $\Delta\MW$ (total)     &   {\bf 84 MeV} &    {\bf 69 MeV}  \\ \hline
\end{tabular}
\end{indented} 
\end{table}

\subsubsection{Direct reconstruction}

During the second period of running in 1996 the \LEP\ experiments
 collected a similar total integrated luminosity
(approximately $40pb^{-1}$) at an energy of 172 $\gev$ to that obtained 
at threshold.
However the future accuracy of the direct reconstruction method for
measuring $\MW$ is more difficult to assess: the analyses are
currently at a relatively preliminary stage, and these data are expected to
comprise a relatively small proportion of those provided by the 
full \LEPTWO\ programme.
 Much more work directed
at reducing the systematic errors will be performed as more data are collected.
Nevertheless, based on existing experience and the Yellow Book studies  
it is possible to make
some educated guesses as to the likely ultimate precision.
As a benchmark, we take the 
current \ALEPH\ and \DELPHI\ (see \cite{JT}) statistical 
and systematic errors{, summarised in Table~\ref{tab:drcurrent},
based on approximately ${\cal L} = 10\pb^{-1}$ per experiment at
$172\gev$. 
Note that the  systematic errors currently range from $60\mev$ to 
$180\mev$ depending on the channel and the experiment.
The major challenge is to reduce these to the ${\cal O}(20 -25\mev)$
level anticipated in \cite{YB}, see elsewhere in this report.
As discussed above,  a \LEP\ beam energy uncertainty of $15\mev$ 
appears to be achievable.
\begin{table}[htb]
\caption{Current statistical and systematic errors (in MeV) on $\MW$
by direct reconstruction from the \ALEPH\ (A) and \DELPHI\ (D)
experiments. The statistical errors are estimated errors.}
\label{tab:drcurrent} 
\lineup 
\smallskip
\begin{indented} 
\item[]\begin{tabular}{|lccc|} \hline
channel & expt. & stat. & sys.\\ \hline
$ \qqln $ &  A & 510 & 60    \\
            &  D & 500 & 94    \\ \hline
$ \qq qq$   & A  & 450 & 180    \\
            & D  & 450 & 75    \\ \hline
\end{tabular}
\end{indented} 
\end{table}

Table~\ref{tab:drexpect} summarises the expected uncertainties
at the three luminosities ${\cal L} = 100, 300, 500\pb^{-1}$ (per experiment).
Also shown are the YB estimates.
Note that the dependence of $\Delta\MW$ on the beam energy 
in the $180 - 200\gev$ range   is expected to be weak \cite{YB}. The
estimates for the $\qqqq$ channel in Table~\ref{tab:drexpect} 
do not include any contributions to the errors from colour reconnection
or Bose Einstein correlations.

\begin{table}[htb]
\caption{Expectations for $\Delta \MW$ (in MeV) combining
all four \LEP\ experiments
for three different  total luminosities. Note that the $\qqqq$ channel
and combined channel 
estimates do not include any additional errors from colour reconnection
or Bose Einstein correlations.}
\label{tab:drexpect} 
\lineup 
\smallskip
\begin{indented} 
\item[]\begin{tabular}{|lcccccc|} \hline
channel & luminosity/expt. & stat. & sys. & \LEP\ & total & YB \\ \hline
$ \qqln $ & $100\pb^{-1}$  & 71 & 21 & 15 & 76 &    \\
            & $300\pb^{-1}$  & 41 & 21 & 15 & 49 &    \\
            & $500\pb^{-1}$  & 32 & 21 & 15 & {\bf 41} & 44 \\ \hline
$ \qqqq$   & $100\pb^{-1}$  & 71 & 24 & 15 & 77 &    \\
            & $300\pb^{-1}$  & 41 & 24 & 15 & 50 &    \\
            & $500\pb^{-1}$  & 32 & 24 & 15 &  {\bf 43} &  45 \\ 
            &                &  &  &  & ($\oplus$col.rec.$\oplus$BE) & \\ \hline
 combined   & $500\pb^{-1}$  &  &  &  &  {\bf  35} & 34 \\ 
            &                &  &  &  & ($\oplus$col.rec.$\oplus$BE) & \\ \hline
\end{tabular}
\end{indented} 
\end{table}

% CJP
Hence the \MW\ results obtained on the 1996 data sample (using both methods)
are broadly in agreement with the YB predictions. The opportunity is available at 
\LEPTWO\ to obtain precision measurements of \MW\ from these two independent methods. 
We conclude that there is every reason to believe that the direct reconstruction
method YB estimate of $\Delta\MW < {\cal O} (50)\mev$ can indeed be achieved. 
 
\section{Summary}
\label{conc}
 After one year of \LEPTWO\ data taking, measurements of the W mass have been
made using two different techniques: from the threshold cross-section 
and from direct reconstruction. The results from these methods are compatible
and the errors are dominated by their statistical components. 
A common theme of the various studies performed at the workshop was to 
ascertain whether the systematic errors were well understood.
That is, are they understood at a level where
one can extrapolate with confidence to the full \LEPTWO\ statistics and
ensure that systematic effects will not dominate?

 The effects of colour recombination and Bose-Einstein statistics on the
reconstructed W mass in the fully hadronic channel were explored in some
detail. These studies supersede those made for the CERN \LEPTWO\ Workshop
and in both cases suggest that the error could well be smaller than
previous studies have indicated; although more work is needed for a
definitive conclusion on these topics. The importance of relating the
predictions of the various models to measurable quantities, such as
multiplicities, was stressed.

 The background from the \QCD\ 
processes $\epem \to (\Zz/\gamma)^* \to \qq\qq, \qqgg$ to $\WW \to 4\;$jets
is large and needs to be well understood. 
One particular problem is the extent to which the precise \LEPONE\ studies
can be extrapolated to  \LEPTWO\ data. A study at the workshop ascertained
that the relevant \WW\ events do populate the ranges of the four-jet angular
variables for which there are discrepancies between data and Monte Carlo 
at \LEPONE\ . An attempt was also made to understand the differences between 
the \PaSh\ \MC\ and  matrix element predictions for angular variables.
Part of these differences was found to be due to
the lack of correct angular correlations  in the $\qq\qq$ 
part of the \PaSh\ \MC s.

 The correction factors in going from the measured \WW\ four-fermion
cross-section to the \CCTHREE\ cross-section, as determined by the \LEP\
experiments, were compared. This was difficult because of the different
ways that these were implemented in practice. General agreement in these
calculations was obtained.

 A detailed investigation was made of many of the systematic effects which
need to be understood to make a precise measurement of \MW\ by direct
reconstruction. These included studies of the problems of finding the
correct pairing in fully hadronic decays and of the benefits and
disadvantages of the various strategies of kinematic fitting and
rescaling techniques. It was found that for these and other 
studies (on biases, linearity of response, distortion of the lineshape
by non-\CCTHREE\ graphs etc.) there was sufficient understanding at present
but that further studies were needed to ensure the full \LEP\ statistics can
be exploited. The \LEP\ energy error, which is a common systematic to all
measurements of \MW\ can probably be determined with sufficient accuracy,
provided the much needed studies show no surprises.

 Finally, a comparison of what can be expected if there was to be a further
threshold run, and a review of the expected precision of the direct 
reconstruction method, were reported. From experience of running
at $161\gev$, it appears that the precision on $\MW$
from the threshold cross section method estimated in the Yellow Book
study can be improved upon.  Overall, provided there are on-going
studies on the systematics, the desired goal 
%of $\Delta\MW \leq \mathcal{O}$(50) MeV, seems achievable.
of $\Delta\MW \leq {\cal O}$(50) MeV, seems achievable.

\end{document}